\documentclass[twocolumn]{aastex63}

\pdfoutput=1

\usepackage{color}
\usepackage{mathtools}
\def\tess{\hbox{$\textit{TESS}$}} 

\def\kepler{\hbox{$Kepler$}}

\def\numflares{\hbox{4}}

\usepackage{verbatim}

\newcommand{\msol}{M_\odot}

\newcommand{\E}[1]{\hspace{-0.01in}\times\hspace{-0.02in}10^{#1}}  
 
\newcommand{\angstrom}{\mbox{\normalfont\AA}}

\newcommand{\ha}{\rm H\alpha}
\defcitealias{Medina2020}{M20}

\accepted{to the Astrophysical Journal}

\shorttitle{}
\shortauthors{Medina et al.}
\graphicspath{{./}{figures/}}

\begin{document}

\title{Galactic Kinematics and Observed Flare Rates of a Volume-Complete Sample of Mid-to-Late M-dwarfs:  Constraints on the History of the Stellar Radiation Environment of Planets Orbiting Low-mass Stars}

\correspondingauthor{Amber Medina}
\email{amber.medina@austin.utexas.edu}

\author[0000-0002-0786-7307]{Amber A. Medina}
\affiliation{Center for Astrophysics | Harvard \& Smithsonian, 60 Garden Street, Cambridge, MA~02138, USA}
\affiliation{Department of Astronomy, The University of Texas at Austin, Austin, TX 78712, USA}

\author[0000-0001-6031-9513]{Jennifer G. Winters}
\affiliation{Center for Astrophysics | Harvard \& Smithsonian, 60 Garden Street, Cambridge, MA~02138, USA}

\author{Jonathan M. Irwin}
\affiliation{Center for Astrophysics | Harvard \& Smithsonian, 60 Garden Street, Cambridge, MA~02138, USA}

\author[0000-0002-9003-484X]{David Charbonneau}
\affiliation{Center for Astrophysics | Harvard \& Smithsonian, 60 Garden Street, Cambridge, MA~02138, USA}

\begin{abstract}

We present a study of the relationship between galactic kinematics, flare rates, chromospheric activity, and rotation periods for a volume-complete, nearly all-sky sample of 219 single stars within 15 parsecs and with masses between 0.1$-$0.3 $\msol$ observed during the primary mission of TESS. We find all stars consistent with a common value of $\alpha$=1.984 $\pm$ 0.019 for the exponent of the flare frequency distribution. Using our measured stellar radial velocities and Gaia astrometry, we determine galactic UVW space motions. We find 64\% of stars are members of the Galactic thin disk, 5\% belong to the thick disk, and for the remaining 31\%, we cannot confidently assign membership to either component. If we assume star formation has been constant in the thin disk for the past 8 Gyr, then based on the fraction that we observe to be active, we estimate the average age at which these stars transition from the saturated to the unsaturated flaring regime to be 2.4 $\pm$ 0.3 Gyr. This is consistent with the ages that we assign from galactic kinematics: We find that stars with Prot $<$ 10 days have an age of 2.0 $\pm$ 1.2 Gyr, stars with 10 $<$ Prot $\leq$ 90 days have an age of 5.6 $\pm$ 2.7 Gyr, and stars with Prot $>$ 90 days have an age of 12.9 $\pm$ 3.5 Gyr. We find that the average age of stars with Prot $<$ 10 days increases with decreasing stellar mass from 0.6 $\pm$ 0.3 Gyr (0.2 - 0.3 $\msol$) to 2.3 $\pm$ 1.3 Gyr (0.1--0.2 $\msol$).

\end{abstract}

\section{Introduction}
Main-sequence stars with convective envelopes display magnetically induced phenomena, such as star spots, coronal x-ray emission, and flares. Stellar flares are believed to be the result of the reconnection of magnetic field lines \citep{Martens1989}. Flares are observed at all wavelengths extending from x-ray to radio. With dedicated space-based photometric missions like \kepler/K2, and the Transiting Exoplanet Survey Satellite (\textit{TESS}),  white-light flares, which are flares that that emit broadband emission extending from the near-ultraviolet to optical wavelengths, have been studied extensively \citep[e.g.][]{Hawley2014,Davenport2014,Davenport2016,Ilin2019,Davenport2019,Gunther2020,Feinstein2020,Medina2020}. 

Previous studies have found that both the flare rate and the total luminosity emitted by all flares summed together for a given star over some length of time decreases with increasing rotation period \citep{Davenport2016,Ilin2019,Davenport2019,Howard2019,Ilin2021,Medina2020}, but this relationship is not simple. Stars show a saturated level of magnetic activity that is independent of the rotation period up to a critical value; beyond this value, stars transition to the unsaturated regime where coronal x-ray emission \citep{wright2011, Wright2018}, ultraviolet emission \citep{France2018}, and chromospheric $\ha$ emission \citep{Douglas2014,Newton2017} decreases rapidly with increasing rotation period or Rossby number, Ro. Rossby number is a mass-independent proxy for rotation that is defined as the rotation period normalized by the convective turnover time of the star. In \citet{Medina2020}, hereafter denoted \citetalias{Medina2020}, we studied a sample of 125 fully convective M dwarfs with masses between 0.1 and 0.3 $\msol$ that reside within 15 parsecs and were observed in the southern ecliptic hemisphere during the first year of the \textit{TESS} mission and found that this relationship extends to optical flare rates as well. \citet{Reiners2022} found that the underlying physical mechanism responsible for observed saturation of magnetic activity can be attributed to saturation of the magnetic dynamo itself. However, the age at which the stars transition from the saturated regime to the unsaturated regime, which could have a substantial impact on the planets orbiting these stars is not firmly established.


\citet{Skumanich1972} provided a model to explain observations that indicated stellar rotation periods increase as a function of age. This means that in theory the age of a star can be inferred if the rotation period is known; this method for age determination is known as Gyrochronology \citep{Barnes2003}. By observing the rotational evolution of stars in clusters with different ages, previous studies \citep[e.g.][]{Barnes2003,Mamajek2008,Meibom2011a,Augueros2018} determined that stars with the same mass converge to a single rotation period value by an age that is dependent on stellar mass. For example, the rotation periods of solar-type stars converge to a single value by the age of 650 Myr, the age of the Hyades cluster, in contrast to the rotation periods of fully convective M-dwarfs which have not converged to a single value by this age. The spread in rotation periods makes age estimation using the technique of gyrochronolgy unreliable for the fully convective M dwarfs that are the subject of our study. However,  galactic kinematics can provide a means to estimate the age of a population of M dwarfs based on their motion through and location in the Galaxy. The Galaxy is believed to be made up of three populations: the thin disk, thick disk, and halo, with each forming at different times throughout the evolution of the Galaxy. The stars in each population also have velocity dispersions that differ from each other due to their birth environments. 


\citet{Newton2016} used pre-Gaia parallaxes and proper motion information and found that most fully convective M dwarfs with rotation periods that exceed 70 days have ages in excess of 5 Gyr, while stars with rotation periods less than 10 days have ages that are less than 2 Gyr.  \citet{Newton2017} showed that these kinematically younger stars also display increased chromospheric $\ha$ emission while their slowly rotating counterparts show little or no $\ha$ emission. \citet{Kiman2019} showed that fully convective M dwarfs with high levels of $\ha$ activity have smaller velocity dispersions in the W component of their galactic velocity.

A better understanding of the relationship between age, rotation, and magnetic activity for fully convective M dwarfs is also needed to enable an understanding of their attendant planets.  Transiting planets around these stars, such as the ones in our sample: GJ 1132b \citep{Berta-Thompson2015, Bonfils2018}, LHS 1140bc \citep{Dittmann2017,Ment2019}, LHS 3844b \citep{Vanderspek2019}, TOI-540b \citep{Ment2021}, may be targeted in the near-term for atmospheric studies with the James Webb Space Telescope and next generation of extremely large telescopes. Although we often find planets orbiting inactive stars that have transitioned to the unsaturated regime, at earlier times these stars resided in the saturated regime and their planets suffered the consequences. It is important that we characterize the duration of the extended radiation environment associated with the saturated regime. It will allow us to reconstruct their stellar history, which is essential to understanding the current atmospheres, or lack thereof of these planets.

In this study we present a relationship between magnetic activity, kinematic age, and rotation for all single, or effectively single, fully convective M dwarfs with masses between 0.1 and 0.3 $\msol$ that reside within 15 parsecs and were observed during the primary mission of \textit{TESS}. We deem a star effectively single if it has a binary companion with a separation greater than 63 arcseconds, which corresponds to the width of three \textit{TESS} pixels.  We build upon the study presented in \citetalias{Medina2020} in which we determined relationships between flare rates, rotation, and chromospheric emission among the subset of these stars in the southern ecliptic hemisphere (observed in Year 1 of the \textit{TESS} Mission) to include the northern ecliptic sample of \textit{TESS} stars observed during Year 2 of the \textit{TESS} Mission.In \S~\ref{sec:sample} we describe the stellar sample. In \S~\ref{sec:spec_obs}, we describe our spectroscopic observations, our measurement of the equivalent widths of activity indicators, and our measurement of the radial velocities and estimates of the galactic space motions. We describe the photometric observations and measurement of new rotation periods from the MEarth project in \S~\ref{sec:mearth_rot} and discuss the rotation period and flare analysis using \textit{TESS} photometry in \S~\ref{sec:tess}.  We present the results in \S~\ref{sec:results} followed by the discussion and conclusion in \S~\ref{sec:DC}.

\section{Stellar Sample}\label{sec:sample}
In this paper we will be discussing two samples of stars. Both samples are subsets of the volume-complete sample of 512 M dwarfs with masses between 0.1--0.3 $\msol$ that reside within 15 parsecs compiled by \citet{Winters2021}. 

The first sample we will discuss, which we refer to as \textit{the northern ecliptic sample} are the stars observed during Year 2 of the \textit{TESS} Mission, and indeed all of these stars fall within the northern ecliptic hemisphere. For these stars we present measurements of the flare rates,  rotation periods, and spectroscopic activity indicators. To compile the northern ecliptic sample, we begin with the \citet{Winters2021} sample and retain only those stars observed in \textit{TESS} sectors 14-26. We used the information provided in \citet{Winters2021} to remove any stars that are known eclipsing binaries, single-lined or doubled-lined spectroscopic binaries, and visual binaries with separations less than the width of three \textit{TESS} pixels (63 arcseconds).  Our northern ecliptic sample comprises 94 stars that are either single, or are separated by their companions by at least 3 \textit{TESS} pixels. \citet{Winters2021} determined the distances to these stars using parallaxes from the second data release of Gaia \citep{GaiaDR22018} and their masses using the K$_s$ magnitude and the relations presented in \citet{Benedict2016}. The uncertainties on the masses range from 4.7\% to 14.0\%. 

The second sample of stars, which we will refer to as \textit{the all-sky sample} is the union of stars from the northern ecliptic sample discussed above and the sample of 125 stars from the southern ecliptic hemisphere observed in \textit{TESS} sectors 1-13 presented in \citetalias{Medina2020}. In total there are 219 effectively single fully convective M dwarfs that reside within 15 parsecs in the all-sky sample. In Table \ref{tab:MT}, we list the names, TIC identifiers, coordinates, masses, distances, proper motions, and \textit{TESS} magnitudes for the all sky sample. Within that Table, we provide additional quantities, including spectroscopic activity indicators, radial velocities, rotation periods, Rossby numbers, flare rates, and spatial velocities in galactic coordinates. These quantities are either derived as described in the following sections, or reproduced from \citetalias{Medina2020}.

\begin{deluxetable*}{lccl}
\tabletypesize{\scriptsize}
\tablecaption{Stellar Properties, Equivalent Widths, Flare Rates, Radial Velocities, and Space Velocities (Table Format) \label{tab:MT}}
\tablehead{ 
\colhead{Column} & 
\colhead{Format} &  
\colhead{Units} & 
\colhead{Description}} 
\startdata 
1 & A22 & ... & Star Name \\
2 & A10 & ... & TIC Identifier \\
3 & F4.4 & hh:mm:ss.s & R.A in hours, minutes, seconds (J2000) \\
4 & F4.4 & dd:mm:ss.s & Declination in degrees, minutes, seconds (J2000) \\ 
5 & F3.2 & $\msol$ & Stellar mass \\
6 & F3.2 & $\msol$ & Uncertainty in stellar mass \\
7 & F3.2 & parsecs & Distance \\
8 & F5.2 & days    & Photometric Rotation Period \\
9 & F3.4 & mag & Semi-amplitude of variability \\
10 & F3.4 & mag & Uncertainty in semi-amplitude  \\
11 & A1  & ... & Reference for Rotation Period \\
12 & F3.4 & ... & Rossby Number (Prot/$\tau$) \\
13 & F3.2 & mag & \tess~Magnitude \\
14 & F3.3 & $\angstrom$ & Equivalent Width of $\ha$ \\
15 & F3.3 & $\angstrom$ & Uncertainty in equivalent width of $\ha$ \\
16 & F3.3 & $\angstrom$ & Equivalent Width of Calcium II at 8542.09 $\angstrom$ \\
17 & F3.3 & $\angstrom$ & Uncertainty in equivalent width of Calcium II at 8542.09 $\angstrom$ \\
18 &  F3.3 & $\angstrom$ & Equivalent Width of Helium I D$_{3}$  \\
19 & F3.3 & $\angstrom$ & Uncertainty in equivalent of  Helium I D$_{3}$ \\
20 & F3.2 & Log Flares day$^{-1}$ & Natural log rate of flares per day at energies greater than 3.16$\times 10^{31}$ ergs \\
21 & F3.2 & Log Flares day$^{-1}$ & Uncertainty in natural log rate of flares per day at energies greater than 3.16$\times 10^{31}$ ergs \\
22 & F3.3 &  mas & Parallax \\
23 & F3.3 &  mas & Parallax Uncertainty \\
24  & F3.3 &  mas~yr$^{-1}$ & Proper Motion RA \\
25  & F3.3 &  mas~yr$^{-1}$ & Proper Motion RA Uncertainty \\
26  & F3.3 &  mas~yr$^{-1}$ & Proper Motion DEC \\
27  & F3.3 &  mas~yr$^{-1}$ & Proper Motion DEC Uncertainty \\
28  & F3.3 &  km s$^{-1}$ & Radial Velocity  \\
29  & F3.3 &  km s$^{-1}$ & Radial Velocity Uncertainty \\
30 & F3.3 &  km s$^{-1}$ & U$_{lsr}$ space motion \\
31 & F3.3 &  km s$^{-1}$ & U$_{lsr}$ space motion uncertainty\\
32 & F3.3 &  km s$^{-1}$ & V$_{lsr}$ space motion  \\
33 & F3.3 &  km s$^{-1}$ & V$_{lsr}$ space motion uncertainty\\
34 & F3.3 &  km s$^{-1}$ & W$_{lsr}$ space motion \\
35 & F3.3 &  km s$^{-1}$ & W$_{lsr}$ space motion uncertainty \\
36 & F3.3 &  km s$^{-1}$ & Total space velocity \\
37 & F3.3 &  km s$^{-1}$ & Total space velocity uncertainty \\
38 & F3.3 & ... & Thin disk membership probability \\
\enddata 
\tablecaption{Full table available in machine-readable form.}
\tablecomments{Upper limits on the Log Flares day$^{-1}$ is denoted by a value of 0.0 for Column 21, the uncertainty on the  Log Flares day$^{-1}$.}
\end{deluxetable*}

\section{Spectroscopic Observations}
\label{sec:spec_obs}
We used the Tillinghast Reflector Echelle Spectrograph (TRES) located on the 1.5 meter Tillinghast Reflector telescope at the Fred Lawrence Whipple Observatory (FLWO) on Mount Hopkins, Arizona to obtain three or more high-resolution spectra for each target in the all-sky sample that had a declination greater than -15 degrees. We obtained spectra from September of 2016 to January of 2021. TRES covers the wavelength range 3900-9100 $\angstrom$ and has a resolution R$\approx$44,000. Exposure times ranged from 120s to 3~$\times$~1200s to reach a signal to noise ratio of 10-40 at 7150 $\angstrom$. We reduced the spectra using the standard TRES pipeline \citep{Buchhave2010}. 

For stars in the all-sky sample with declinations below -15 degrees we used the the HIgh ResolutiON (CHIRON) spectrogragh \citep{Tokovinin2013} located on the Cerro Tololo Inter-American Observatory (CTIO) SMARTS 1.5 meter Telescope  \citepalias[see][for details]{Medina2020}. CHIRON has wavelength coverage that extends from 4100--8700 $\angstrom$. We used the image slicer mode to achieve a resolution of R$\approx$80,000. We gathered spectra from December of 2017 to March of 2021. We reached a signal to noise of 3-15 per pixel at the wavelength of 7150 $\angstrom$ using exposure times that ranged from 180 seconds to 3 $\times$ 1800 seconds. We used the standard CHIRON reduction pipeline to extract each spectrum \citep{Tokovinin2013,Paredes2021}.

\subsection{Spectroscopic Activity Indicators}

In \citetalias{Medina2020}, we presented, for all stars in the southern ecliptic portion of the all-sky sample, the equivalent widths (EWs) of the spectroscopic magnetic activity indicators Helium I D3 at 5875.6 $\angstrom$, $\ha$ at 6562.8 $\angstrom$, and one of the three Calcium infrared triplet lines, 8542.09 $\angstrom$. We did not report measurements of the other two members of the Calcium infrared triplet as they fell outside of the CHIRON spectral format. We reproduce these values in Table \ref{tab:MT} for completeness.

We used FLWO/TRES to measure the EWs for these same three spectral features for all stars in the northern ecliptic portion of the all-sky sample. We follow the methods described in \citetalias{Medina2020}: we use the wavelength ranges for the feature and surrounding continuum regions as listed in Table \ref{tab:EW}. We take negative EW values to denote emission and state the maximum EW value (least amount of emission) of our multi-epoch observations as we believe this value provides a better representation of the quiescent level of emission. We report these values in Table \ref{tab:MT}.




\begin{center}
\begin{deluxetable}{lccc}
\tablecaption{Regions used for Measurement of Equivalent Widths \label{tab:EW}}
\tablehead{ 
\colhead{Feature} & 
\colhead{F$_{C1}$} & 
\colhead{F($\lambda$)} &   
\colhead{F$_{C2}$}  \\
& 
\colhead{$\angstrom$} & 
\colhead{$\angstrom$} &
\colhead{$\angstrom$}}  
\startdata
$\ha$         &  6554.1 $-$ 6559.1 & 6560.3 $-$ 6565.3 & 6566.5 $-$ 6570.5  \\
He I D$_3$  & 5870.0 $-$ 5873.0 & 5874.6 $-$ 5876.6   & 5877.6 $-$ 5880.6 \\
Calcium II    & 8537.0 $-$ 8540.0 &   8541.3 $-$ 8542.8  & 8560.0 $-$ 8580.0 \\
\enddata

\end{deluxetable}
\end{center}

\subsection{Radial Velocities and Galactic Kinematics}\label{sec:gal_kin}


We used cross-correlation methods similar to those described in \cite{Kurtz(1998)} to measure radial velocities for every star in the all-sky sample. For our
template, we used a high-SNR spectrum of Barnard's Star, observed with
TRES on UT 2018 July 19 and with CHIRON on UT 2018 April 22. Barnard's
Star is a slowly rotating \citep[130.4 days,][]{Benedict(1998)} M4.0
dwarf \citep{Kirkpatrick(1991)} for which we adopted a Barycentric
radial velocity of $-110.3\pm0.5\ {\rm km\ s^{-1} }$, derived from
presently unpublished CfA Digital Speedometer \citep{Latham(2002)}
measurements taken over 17 years.  We saw negligible rotational
broadening in our Barnard's Star template, in agreement with the $v
\sin i$ of 0.07 km s$^{-1}$ ~expected from its long photometric rotation
period. This is also consistent with the $v \sin i$ upper limit of $2
~{\rm km\ s^{-1} }$ reported by \citet{Reiners(2018)}.

We used methods similar to those described in
\citet{Winters(2018),Winters2020}, which used only a single order for each spectrograph (order 41 for TRES and order 44 for CHIRON). Here we used the weighted average of the RVs derived from from each of 6 orders for each spectrograph (orders and central wavelengths, 36 (6461.1 $\angstrom$), 38 (6714.6 $\angstrom$), 39 (6820.7 $\angstrom$), 41 (7115.0 $\angstrom$), 43 (7444.8 $\angstrom$), and 45(7783.3 $\angstrom$) for TRES, and orders 36 (6436.5 $\angstrom$), 37 ( 6510.5 $\angstrom$), 39 (6663.7 $\angstrom$), 40 (6743.0 $\angstrom$), 44 (7080.1 $\angstrom$) and 51 (7759.0 $\angstrom$) for CHIRON). We selected these orders from the red portion of the format (so that there was sufficient SNR), avoided the $\ha$ feature, and were reasonably free of telluric lines, such that the peaks of the correlation functions generally exceeded 0.5 in well-exposed spectra.

We calculated the RVs in each order, with uncertainties calculated using
the methods described in \citet{Bouchy(2001)}. The RVs in each order were then combined using the inverse variance weighted mean, with the uncertainty being the standard error in the inverse weighted mean. We repeated this for each epoch. The final reported RV is the weighted mean from all available epochs. The final reported uncertainty is the weighted mean uncertainty from all available epochs scaled by 1.65 and 1.8 for TRES and CHIRON respectively to reach agreement with the observed velocity scatter and then added in quadrature with the 0.5 km s$^{-1}$ RV uncertainty of the Barnard's star template spectrum; this is the RV uncertainty we present in Table \ref{tab:MT}. For six of our targets that were faint and/or rapidly rotating, the spectra had extremely low SNR (SNR $\leq$ 2), we use only the echelle order with the highest signal to noise and hence most reliable RVs to compute the cross-correlation; for TRES this was order 41 and for CHIRON this was order 44.

In the cases where there is measurable rotational broadening, we
measured the $v \sin i$ per target spectrum as follows. We performed two nested grid searches for the maximum peak correlation, $h$, over a $v \sin i$ range of $0-100$ km s$^{-1}$ sampled at 1 km s$^{-1}$. We calculated a peak correlation against a broadened template spectrum. The template spectrum is broadened by convolving the template with a standard limb darkened rotational broadening kernel \citep{Gray2005} . We then used parabolic interpolation of the quantity $h^2$ to obtain a more precise estimate of the peak. To further refine the estimate, we repeated this procedure $\pm$ 1 km s$^{-1}$ about the best value from the first grid sampled at 0.1 km s$^{-1}$. We again used parabolic interpolation to obtain the final value of $v \sin i$ from the 0.1 km s$^{-1}$ grid results. We then averaged the $v \sin i$ per object and used this value to apply rotational broadening to the template spectrum when deriving the RVs.


In order to compute the spatial velocities of the all-sky sample, we use proper motions and parallaxes as well as their associated uncertainties from the second data release of Gaia \citep{GaiaDR22018}. For one star 2MASS J0943-3833, our RV analysis was inconclusive as we could not find a peak in the CCF and the spectrum was very low signal to noise. We remove this star from further kinematic analysis. For the remaining 218 stars, we used the radial velocities, proper motions, and parallaxes to compute the U,V, and W space motions for each star following the definition of \citet{Johnson1987}, where U describes the velocity towards the galactic center, V describes the velocity with respect to the rotation of the galaxy where positive is in the direction of galactic rotation, and W describes the vertical velocity with positive denoting in the direction of the North Galactic pole. We measure the velocities relative to the barycenter of the solar system.  The uncertainties in U, V, and W are calculated using the method described in \citet{Johnson1987}. The dominant source of error in the calculation of the UVW space motion comes from our measured radial velocities. In Table \ref{tab:MT}, we list the proper motions, parallaxes, radial velocities, and UVW space motions measured relative to the local standard of rest as well as derived quantities defined in the following sections including the total space velocity, which is the magnitude of the combined U$_{lsr}$, V$_{lsr}$, and W$_{lsr}$ velocity components, and the probability ratio of each star residing in the thick disk versus the thin disk. To convert the component  velocities to LSR, we add the LSR component velocities measured by \citet{Schonrich2010} of 11.10 km~s$^{-1}$, 12.24 km~s$^{-1}$, 7.25 km~s$^{-1}$ for U$_{lsr}$, V$_{lsr}$, W$_{lsr}$.


\section{New Photometric Rotation Periods From MEarth}\label{sec:mearth_rot}
Inhomogeneously distributed stellar spots rotating into and out of view can produce a photometric modulation that can be used to measure stellar rotation periods. The MEarth North and South Observatories each consists of eight 40-inch telescopes located atop Mt. Hopkins in Arizona and CTIO in Chile \citep{Nutzman2008,Irwin2015} and have been used extensively since 2008 and 2014 respectively to determine photometric rotation periods for fully convective M dwarfs. For the 219 stars in our all-sky sample, 67 had previously published rotation periods measured from MEarth data \citep{Newton2016,Newton2018,Diez2019,Morales2019,Vanderspek2019}. In \citetalias{Medina2020}, we presented 18 additional rotation periods; 17 of these were from MEarth-South and one was from MEarth-North. Here we followed the methods described in \citet{Newton2016, Newton2018} and used in \citetalias{Medina2020} to measure an additional eight rotation periods, all from MEarth-North. We list the new and previously determined MEarth rotation periods and semi-amplitudes of variability in Table \ref{tab:MT}. In Table \ref{tab:MT} we also list two additional rotation period measured from other ground-based data sets. \citet{Suarez2016} used data from the All-Sky Automated Survey to determine the rotation period for GJ 54.1 (P = 69.2 days) and \citet{Diez2019} used data from the SuperWASP telescope to determine the rotation period of LEP 2211+4059 (P = 30.0 days). We independently confirm the rotation period of GJ 54.1 using MEarth data, but do not currently have sufficient data of LEP 2211+4059 to confirm its rotation period.
 
\section{\textit{TESS} Photometry}\label{sec:tess}
We analyzed the \textit{TESS} light curves of the 94 stars in our northern ecliptic sample (the data from year 2 of \textit{TESS}). We largely follow the methods described in \citetalias{Medina2020}. We summarize the methods below, and we refer the reader to that paper for the full description. We used the pre-search data conditioning simple aperture photometry (PDCSAP). We used all data with a quality flag equal to zero or 512. Quality flag 512 denotes a point that is an impulsive outlier.  \citet{Feinstein2020} and \citetalias{Medina2020} found that points during a flare can be assigned a quality flag of 512, and so we retain them.

\subsection{Photometric Rotation Periods with \textit{TESS}}
\begin{figure}[ht]
\includegraphics[scale=.5,angle=0]{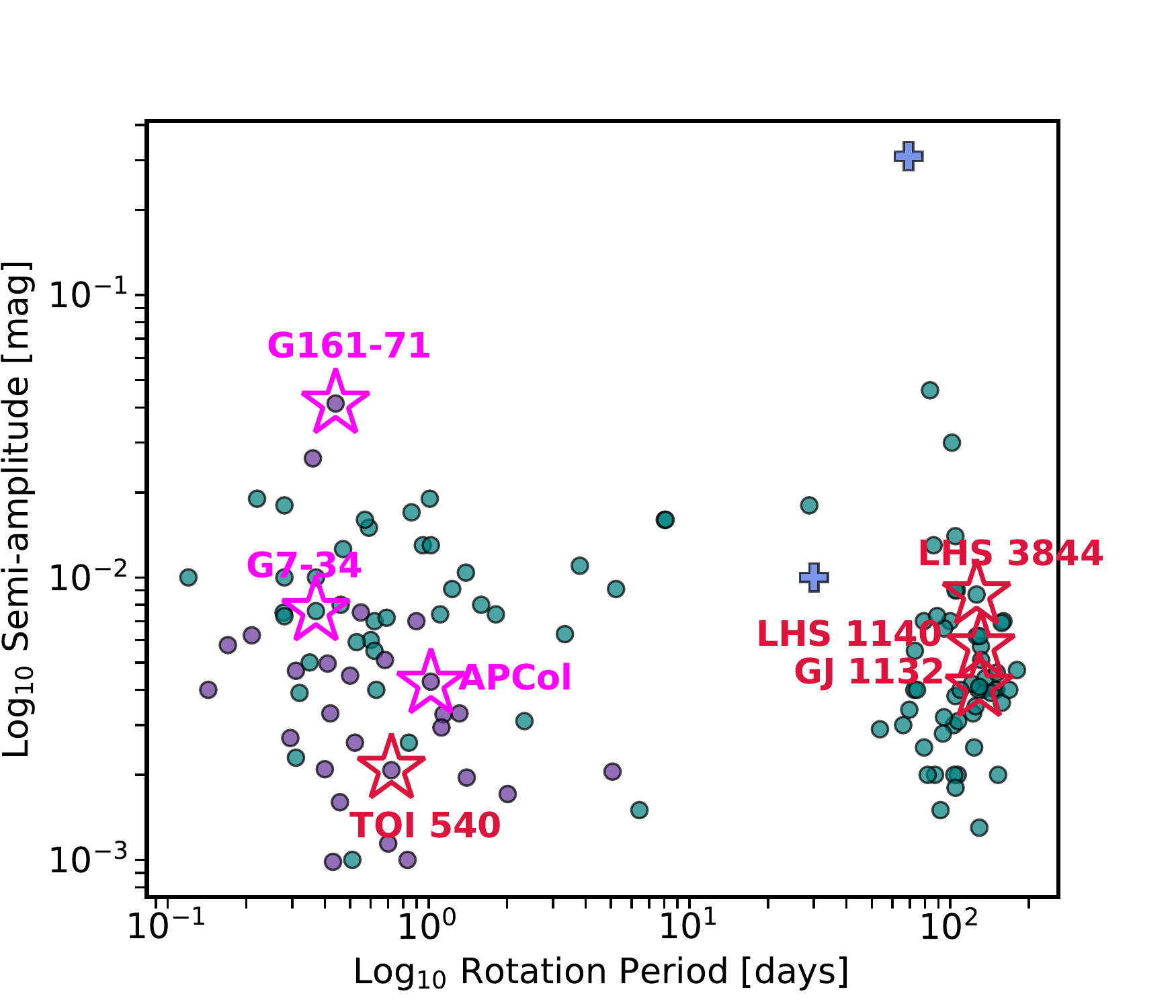}
\caption{The semi-amplitude of variability as a function of rotation period. The teal circles show rotation periods determined using MEarth data presented in \citet{Newton2016}, \citetalias{Medina2020}, and this work. The purple circles show the rotation periods measured in this work and \citetalias{Medina2020} using \textit{TESS} data. The two blue plus symbols represent stars that did not use MEarth or \textit{TESS} data for the determination of the rotation periods of Gl 54.1 and LEP 2211+4059.  Red stars denote known transiting planet hosts. Magenta stars denote members of young moving groups discussed in section \ref{sec:ymg}.  We see no relationship between the rotation period and the semi-amplitude of variability. \label{fig:prot}}
\end{figure}

We found seven new rotation periods with \textit{TESS} ranging from 0.29 to 1.31 days. We used the methods described in \citet{Irwin2011,Newton2016,Newton2018,Medina2020} to measure these rotation periods. We also present the rotation period for one star, WT 887, from \citetalias{Medina2020}. WT 887 has the shortest measured rotation period (P=3.4h) of any star in the all-sky sample. The $v \sin i$ = 26 km s$^{-1}$ we measure for this object from our CHIRON spectra confirms this rapid rotation.
For the 24 stars with previously published rotation periods less than 14 days in the northern ecliptic sample that were initially presented in \citet{Newton2016},  we measure their rotation periods using the \textit{TESS} data to check for consistency. We find consistency between the rotation period we measure with the \textit{TESS} data and the previously published period for 23 stars. For one star, LP 255-11, P$_{rot}$ = 8.04 days \citep{Diez2019} the period determination using the \textit{TESS} data is inconclusive. We do however confirm its rotation period using our own analysis of the MEarth North data.  We confirm the previous findings presented in \citet{Newton2016} and \citetalias{Medina2020} that the semi-amplitude of variability is not correlated with rotation period. In Figure \ref{fig:prot}, we show the semi-amplitude of variability as a function of rotation period for the 120  stars among the 219 in our all-sky sample with measured rotation periods from \textit{TESS}, MEarth or other ground-based efforts. We present the new rotation periods as well as the semi-amplitudes of variability in Table \ref{tab:MT}. 

\subsection{Stellar Flares with \textit{TESS}}\label{sec:flare_alg}
To search for flares we used the \textit{TESS} two-minute cadence PDCSAP light curves.  We used the same methodology that was described in \citetalias{Medina2020}, which we summarize here for completeness. We removed stellar and instrumental correlated variability using the python package {\texttt {exoplanet}}  \citep{exoplanet:exoplanet,Foreman-Mackey(2017)} Gaussian process (GP) that took the functional form of two simple harmonic oscillators to model the PDCSAP light curve. We then subtract the GP model from the data and search the residuals for at least three consecutive three sigma outliers. Data that meet these criteria are deemed a stellar flare.

\subsubsection{Flare Energies}
We determined the energy of each detected flare following the methods described in \citetalias{Medina2020}, which we summarize here. We multiply the luminosity of the star in the \textit{TESS} bandpass by the equivalent duration (ED) of the flare, which is measured in seconds. Physically the ED is defined as the time that it takes for the star in a non-flaring quiescent state to emit the same energy that was radiated by the flare. We determined the ED by integrating the area (i.e. summing the values) during the flare in the light curve after subtracting the Gaussian process model. As in \citetalias{Medina2020}, we assign a 5\% uncertainty to our flare energies. This uncertainty arises because of the uncertainty in where the flare ends (see \citetalias{Medina2020} for details on how the uncertainty is determined).

In total we find 1959 flare events in the northern ecliptic sample. These events are a mixture of classical single peaked flare morphologies and multi-component morphologies. In this study we were interested in the total energy of each flare event and thus do not differentiate between complex and classical flare morphologies because we simply integrate the total area under the light curve. We found that 72 of the 94 stars in our northern ecliptic sample flare at least once during sectors 14-26. We provide a catalog of the Barycentric Julian date at which the peak of the flare occurred, flare energies, amplitudes, durations, and equivalent durations for the stars in the northern ecliptic sample in Table \ref{tab:flares}.

\subsubsection{Flare Completeness Correction}\label{sec:flare_C}
We computed a completeness function for each star to understand over what energy range of flares our algorithm is sensitive. We followed the methods presented in \citetalias{Medina2020}, which we summarize here for completeness. We injected flares at 30 logarithmically spaced energies ranging from 10$^{28}$ to 10$^{33}$ ergs. We used the flare template described in \citet{Davenport2014} to inject a total of 100 flares at each energy into each light curve by injecting 10 flares into a given light curve 10 times. We assigned each flare a random start time, but we took care to not inject a flare at a time where a real flare was located. The flare template is described by the time of peak flux, the amplitude, and the full width at half maximum of the flare, which is a proxy for the flare duration. As discussed in \citetalias{Medina2020}, the \citet{Davenport2014} template does not describe all flare morphologies. For the purposes of this study however, we are concerned with only flare energies. Thus as long as the flaring event shows at least three positive, three sigma outliers our algorithm will detect it, regardless of its morphology. We then determined the fraction of flares recovered for each energy, which we model using the error function. 

\begin{figure*}
\includegraphics[scale=.7,angle=0]{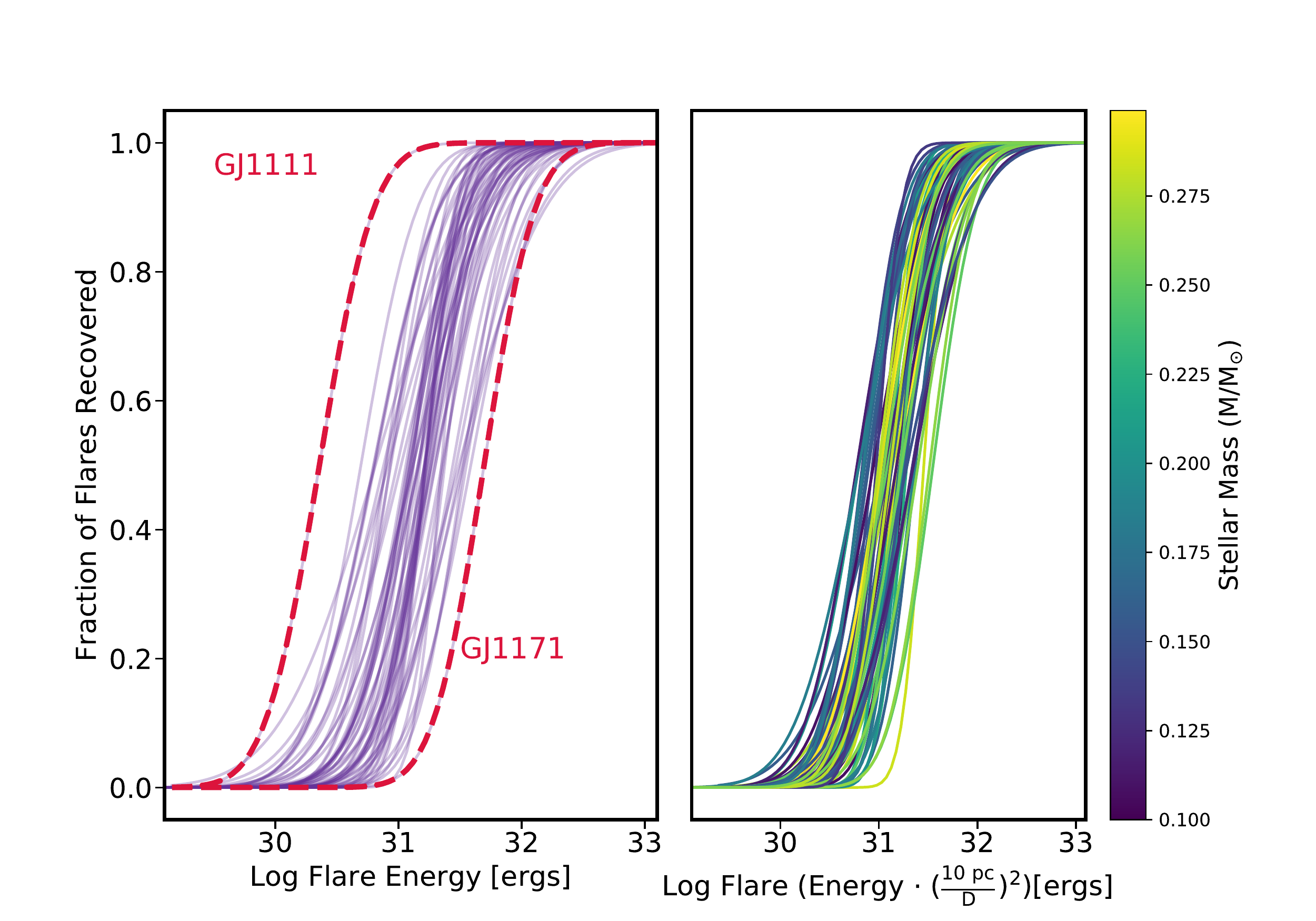}
\caption{Left Panel: The fraction of flares recovered as a function of energy for each star in our northern ecliptic sample. Dashed red lines show the completeness function for the closest star in our northern ecliptic sample, GJ 1111 at 3.58 parsecs (left dashed line) and the for the farthest star in our northern ecliptic sample, GJ 1171 at 14.98 parsecs (right dashed line). The right panel shows the completeness function for each star in our northern ecliptic sample scaled to a common distance of 10 parsecs to account for the inverse square law. After correcting for distance, we see that stellar mass, a proxy for luminosity, is a secondary factor: We are less sensitive to flares of a given energy around stars of greater mass due to a smaller contrast between the flare and the intrinsic stellar luminosity. We use these curves to correct for this changing sensitivity on a star-by-star basis.\label{fig:CF}}
\end{figure*}

We show the completeness function for each star in the northern ecliptic sample in Figure \ref{fig:CF} as well as the completeness function where the results have been scaled to account for the inverse square law to a common distance of 10 parsecs. Figure \ref{fig:CF} demonstrates that distance is the dominant variable in setting flare detection sensitivity, followed by stellar mass.  At the 50\% completeness level, which is the energy at which our algorithm is able to detect at least 50\% of injected flares, our algorithm is able to detect flares with energies above Log$_{10}$(E/ergs) = 30.35 on GJ 1111 (distance = 3.58 parsecs). This is in comparison to the 50\% completeness energy of Log$_{10}$(E/ergs) = 31.46 for GJ 1171, which is the most distant star in our sample (distance = 14.98 parsecs). We use \textit{log} to denote natural log; log base 10 is denoted as \textit{log$_{10}$}.

\begin{deluxetable*}{lccl}
\tabletypesize{\scriptsize}
\tablecaption{Catalog of Stellar Flares in the Northern Ecliptic Sample \label{tab:flares} (Table Format)}
\tablehead{ 
\colhead{Column} & 
\colhead{Format} &  
\colhead{Units} & 
\colhead{Description}}
\startdata 
1 & A22 & ... & Star Name \\
2 & A10 & ... & TIC Identifier \\
3 & F2.2 & mag & \tess~Magnitude \\
4 & F7.4 & BJD & Time of Peak Flux \\ 
5 & F3.2 & cts s$^{-1}$ & Flare Amplitude \\
6 & F4.1 & seconds & Flare Duration \\
7 & F2.2 & seconds  & Equivalent Duration \\
8 & E1.2  & ergs & Flare Energy in the \tess~Bandpass \\
\enddata 
\tablecaption{Full table available in machine-readable form.}
\end{deluxetable*}

\subsubsection{Flare Frequency Distribution}

\citet{Lacy1976} showed that the frequency of flares as a function of energy (known as the flare frequency distribution (FFD)) can be described by,

\begin{equation}\label{eq:FFD}
    N(E)dE = \Omega~E^{-\alpha}dE
\end{equation}

\noindent where $\Omega$ is a normalization constant and $\alpha$ is power-law exponent of the FFD.

In \citetalias{Medina2020}, we found that the fully convective stars in our southern elliptic sample shared the same value of $\alpha = 1.98 \pm 0.02$. This value is consistent with, but more precisely determined than previous flare studies of  much smaller samples of fully convective M dwarfs \citep{Hawley2014,Lurie2015, Silverberg2016}, or those that used data with a much longer cadence of 30 minutes \citep{Ilin2019}.

We determined $\alpha$ for the all-sly sample following the methods described in \citetalias{Medina2020}: first, we used Markov Chain Monte Carlo methods and the python package \textit{emcee} \citep{Foreman2013} to determine $\alpha$ for 39 stars that showed five or more flares during \textit{TESS} observations of sectors 14-26. We then combined these with the 38 values from \citetalias{Medina2020} finding the error weighted mean for the all-sky sample to be $\alpha$ =  1.984 $\pm$ 0.019. For the remainder of this paper, we will present results from the analysis of this combined sample, the all-sky sample.



\begin{figure}[ht]
\includegraphics[scale=.55,angle=0]{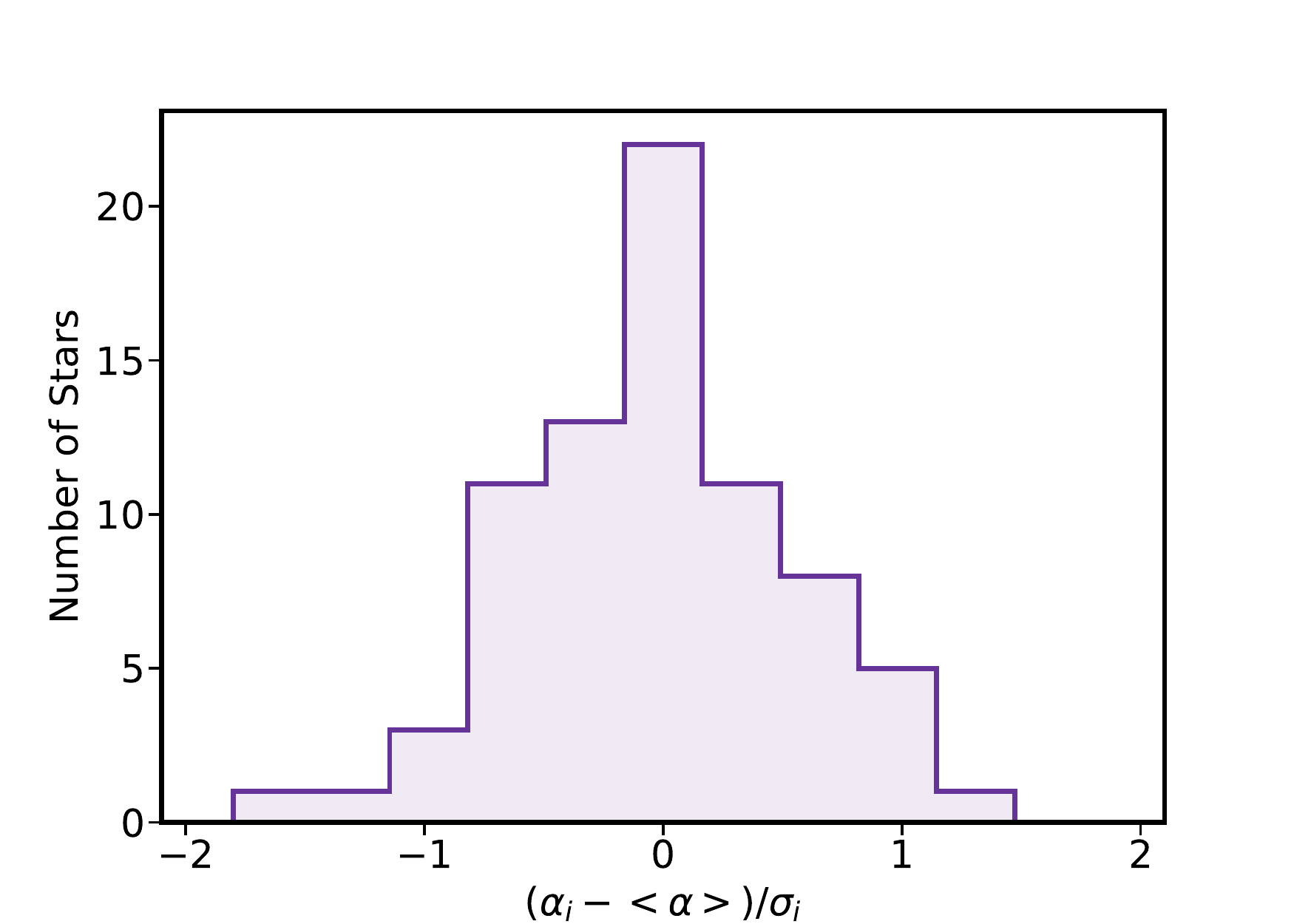}
\caption{Distribution of the Z-scores ($\alpha_i - <\alpha>$)/$\sigma_i$ for fitted values of $\alpha$,  the power-law exponent of the flare frequency distribution for the stars n the all-sky sample for which we detected at least five flares, where $\alpha_i$ are the best fit values of the exponent and $\sigma_i$ are the uncertainties on the exponent estimates and $< \alpha >$ is the error weighted mean of all stars for which we measured $\alpha$. \label{fig:alpha_hist}}
\end{figure}

Figure \ref{fig:alpha_hist} shows the Z-score, which is the difference between the individual $\alpha_i$ and the error-weighted average value $\alpha$, divided by the individual uncertainties $\sigma_i$.  This figure shows that all stars are consistent with a single value of $\alpha$ = 1.984 $\pm$ 0.019.

We then assumed this value for all stars, and proceeded to determine either a flare rate (for stars that had at least one flare), or an upper limit, as described in \citetalias{Medina2020}. We chose to evaluate the flare rate at an energy of 3.16 $\times$ 10$^{31}$ ergs or log$_{10}$(E/ergs) = 31.5 in the \textit{TESS} bandpass. As in \citetalias{Medina2020}, this is the energy at which all stars in our sample are 50\% complete or above i.e. the energy at which our algorithm is able to detect at least 50\% of the flares we inject in the light curve of the star for which we are the least sensitive to flares in the sample. We denote as R$_{31.5}$ the number of flares per day at or above an energy of 3.16 $\times$ 10$^{31}$ ergs. We list the log R$_{31.5}$ values and their uncertainties in Table \ref{tab:MT}. We emphasize that log R$_{31.5}$ is the natural log of the flare rate not log base ten.

\section{Results}\label{sec:results}

\begin{figure*}
\includegraphics[scale=.6,angle=0]{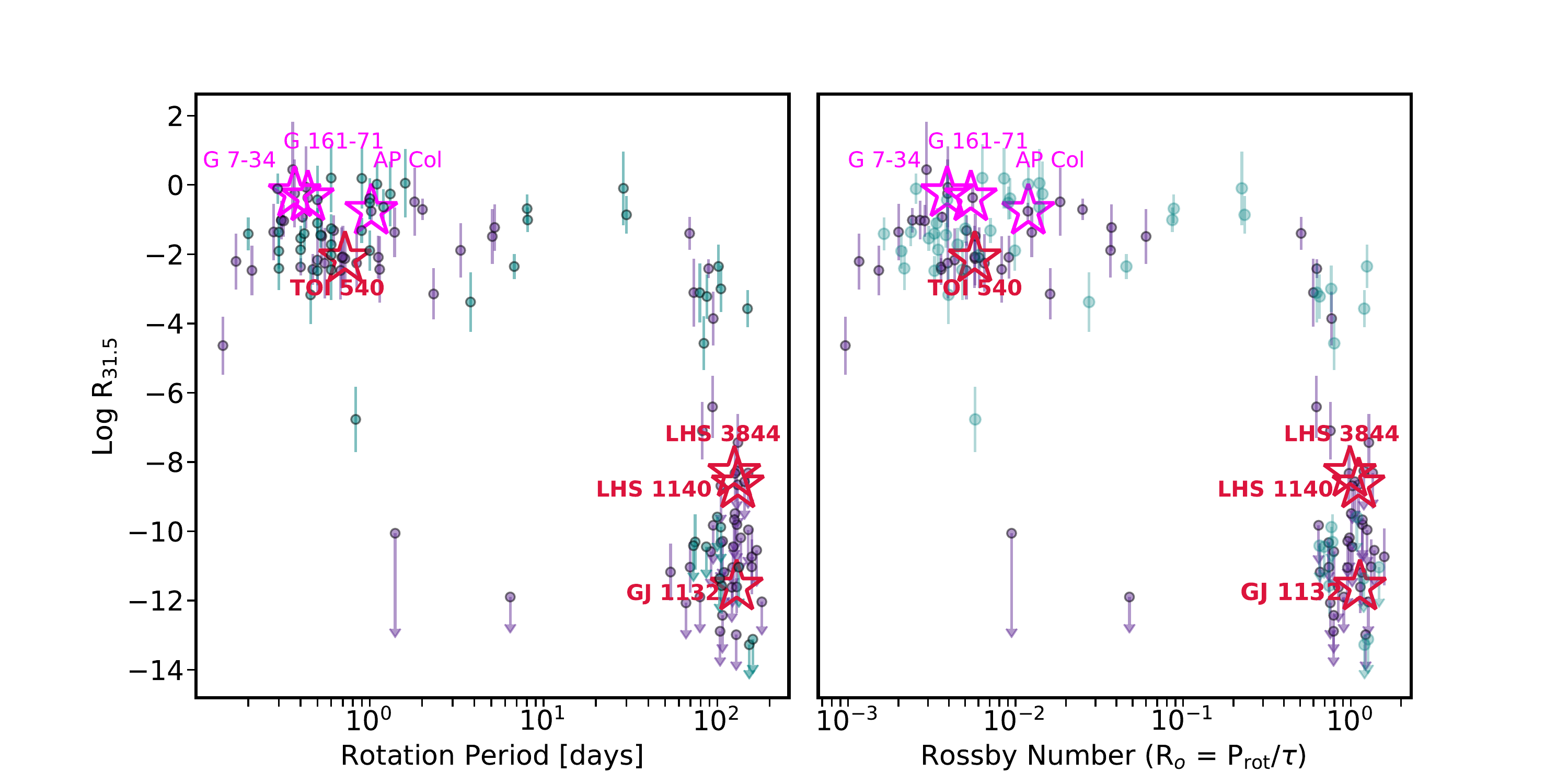}
\caption{The left panel shows the log number of flares per day above our completeness threshold of E = 3.16$\times 10 ^{31}$ ergs as a function of the stellar rotation period. The right panel shows the same flare rate as a function of Rossby number. Purple points are stars observed by \textit{TESS} in the southern ecliptic hemisphere presented in \citetalias{Medina2020} and teal points are stars observed by \textit{TESS} in the northern ecliptic hemisphere. Red stars denote known transiting planet hosts. Magenta stars denote members of young moving groups discussed in section \ref{sec:ymg}. \label{fig:prot_fl}}
\end{figure*}


\subsection{Flare Rate and Stellar Rotation Period}\label{sec:flare_rate_prot}
Rossby number is dimensionless proxy for stellar rotation that is defined as Ro = P$_{rot}$/$\tau$, where P$_{rot}$ is the stellar rotation period and $\tau$, is the mass dependent convective turnover time.
Rossby number is dimensionless quantity that serves as a proxy for stellar rotation. It is defined as Ro = P$_{rot}$/$\tau$, where P$_{rot}$ is the stellar rotation period and $\tau$, is the mass-dependent convective turnover time. The convective turnover time has not been measured directly for any star other than the Sun. We used empirical relations from \citet{Wright2018} to compute the convective turnover time for each star. In Figure \ref{fig:prot_fl}, we show the flare rate as a function of rotation period and Rossby number for the 122 stars in the all-sky sample with a measured rotation period. We found that there are two different regimes of flaring behavior. Stars that have Rossby numbers less than 0.5 show a saturated flare rate equal to log R$_{31.5}$ = -1.32 $\pm$ 0.06,
whereas stars with Rossby numbers greater than 0.5 displayed flare rates that could decrease six orders of magnitude or more (since, for some stars, we only have an upper limit on the flare rate). From Figure \ref{fig:prot_fl}, we note that stars in the saturated regime appear to show an intrinsic scatter of 0.79 in the value of log R$_{31.5}$ about the mean of -1.32.


Instead of a sharp drop in the flare rate for Rossby numbers above 0.5, there exists a region of Rossby numbers above 0.5 for which stars with the same Rossby number can show either a saturated or unsaturated flare rate. This dichotomy in flare rates for a given Rossby number is reminiscent of a finding by \citet{Morin2010} concerning a dichotomy in magnetic field topologies of fully convective M dwarfs. They find that stars with non-axisymmetric fields with quadrupolar and octupolar components have a weak global magnetic field compared to stars of the same rotation period with a dipolar asymmetric fields and a strong global magnetic field. However, by construction, due to the limitations of Zeeman Doppler Imaging, the technique used to measure magnetic field topologies, the stars in the \citet{Morin2010} sample were rapidly rotating, with rotation periods less than 1.5 days. The stars we observe in this region all have rotation periods significantly longer that this. \citet{Brown2014} provides a possible explanation by proposing two different regimes of angular momentum loss separated by an abrupt transition.  At young ages, the stellar dynamo, which is believed to be responsible for the generation of a magnetic field, is weakly coupled to stellar wind making angular momentum losses minimal. At later times, the dynamo abruptly transitions to be strongly coupled to the stellar wind leading to a time of greater angular momentum loss and the star spins down. Perhaps in this region above Ro = 0.5,  we are observing a mix of stars that are in the process of this abrupt transition.


We find that there are four exceptions to this relationship between rotation and flaring: GJ 1187, WT 887, 
SCR~J1855-6914 and 2MASS~J23303802-8455189. These stars have a rotation periods of 0.83, 0.143, 1.40, and 6.43 days and log R$_{31.5}$ = -6.77 $\pm$ 0.94,  -4.64 $\pm$ 0.84, and upper limits of -10.06, -11.9, respectively. All of these stars also show $\ha$ in emission.  We postulate a few reasons for these anomalously low flare rates. One is that these stars may have undetected binary companions, which may lead to incorrect mass estimates erroneously placing them in our sample or perhaps that the flares and rotational modulation are coming from different members of the system. Using a bisector analysis of the spectral line profiles of these four stars (E. Pass private communication) there is no line variation as a function of time that would indicate of a companion that has a flux ratio of at least 0.4 and less than 0.8  and an RV offset of at least 3.0 and less than 8.0 km s$^{-1}$  (E. Pass private communication). This analysis does not rule out equal mass binaries or ones with an RV offset less than 3.0 km s$^{-1}$. However it is unlikely that GJ 1187, WT 887, and SCR-J1855-6914 are unresolved equal mass binaries because they would be over-luminous and we find that their photometric distances and trigonometric distances are in agreement. For 2MASS~J23303802-8455189 the photometric distance is greater than the trigonometric distance indicating this star is actually underluminous. Furthermore, we do not believe any of these stars are astrometric binaries as the Gaia astrometric noise, which can be used as indication of an astrometric binary if the value is greater than two, is less than one for all of these stars. We would like to note that WT 887 is the most rapidly rotating star in our sample with a rotation period of 0.11 days and thus may be demonstrating the phenomenon of super-saturation, wherein extremely rapidly rotating stars show reduced levels of magnetic activity \citep{Doyle2022}.

If a binary companion is not involved, it could be the case that there is some aspect of the evolution of the star that is suppressing magnetic activity. To investigate this, we explored the location of these stars on color magnitude diagrams (CMD). We obtained colors for our sample of stars from Gaia DR2. We find, when examining the absolute K band magnitude as a function of G-K, that WT 887 and SCR~J1855-6914 are some of the lowest mass stars in the sample. This same CMD shows that GJ 1187 and 2MASS~J23303802-845518 have relatively blue colors and thus, may be metal poor which could effect the rotational evolution of the star and thus its activity \citep{Amard2020}.  In any case, we do not include GJ 1187, WT 887, SCR~J1855-6914, or 2MASS~J23303802-8455189 in the calculation of the saturated value of log R$_{31.5}$ = -1.32 $\pm$ 0.06.

\subsection{Flare Rate and Spectroscopic Activity Indicators}
\begin{figure*}
\includegraphics[scale=.45,angle=0]{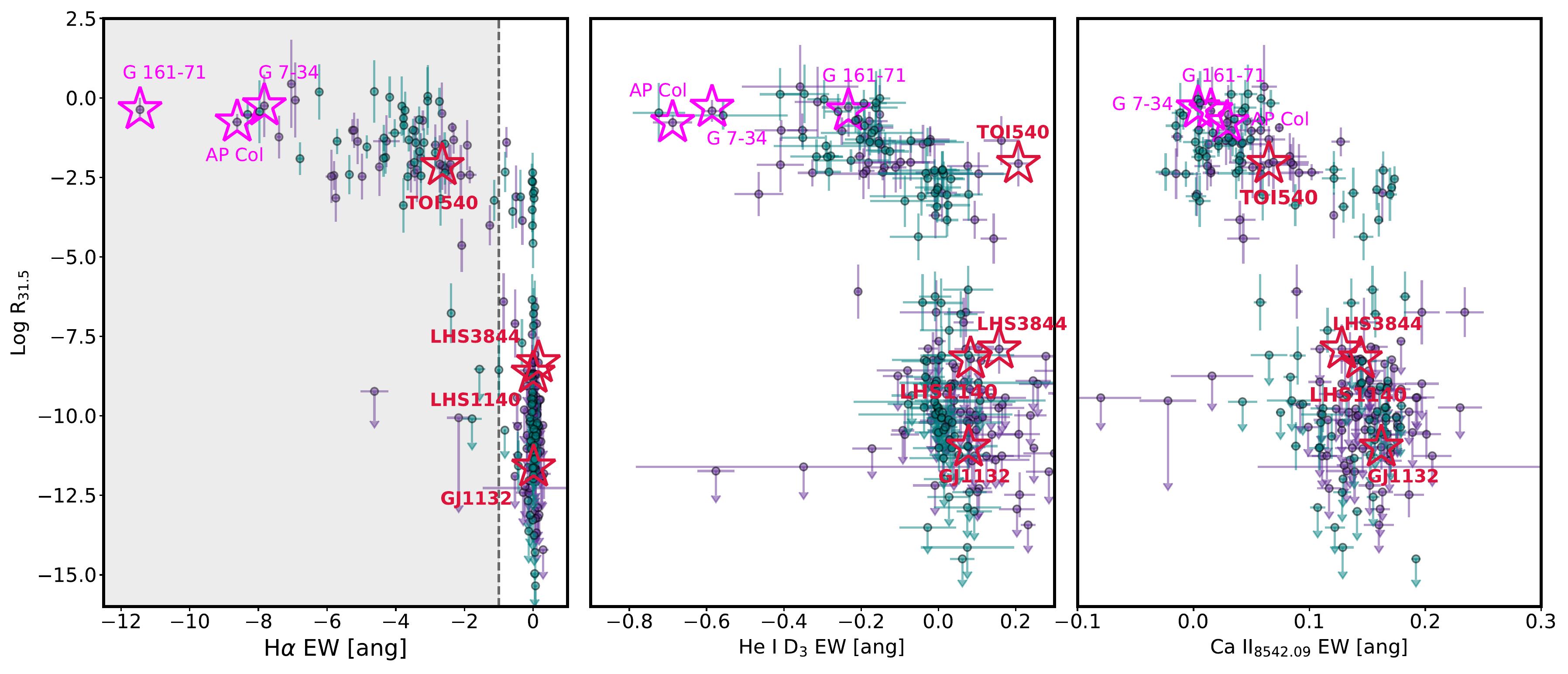}
\caption{The natural log rate of flares per day with energies above E = 3.16$\times$ 10$^{31}$ ergs as a function of measured equivalent widths of $\ha$ (left),  He I D$_3$ (middle), and the calcium infrared triplet line at 8542.1 $\angstrom$ (right). The shaded region in the $\ha$ panel (left) represents stars with $\ha$ EW $<$ -1.0 $\angstrom$, which we classify as active. Upper limits are shown as downward pointing arrows. Purple points are stars observed by \textit{TESS} in the southern ecliptic hemisphere presented in \citetalias{Medina2020} and teal points are stars observed by \textit{TESS} in the northern ecliptic hemisphere. Red stars denote known transiting planet hosts. Magenta stars denote members of young moving groups discussed in section \ref{sec:ymg}. \label{fig:EWs}}
\end{figure*}

In this section we explore the relationship between two magnetically induced phenomena; chromospheric emission and stellar flares. In Figure \ref{fig:EWs}, we show the flare rate as a function of the equivalent widths of $\ha$, He I D$_3$, and one of the calcium infrared triplet lines at 8542.09 $\angstrom$. As in \citetalias{Medina2020}, we find both a saturated and an unsaturated regime, and in all three cases there exists a range of EWs in which the saturated and unsaturated populations overlap:  for $\ha$ this occurs between EWs of -1.0 and 0.0 $\angstrom$. We find $\ha$ to be the most well-separated where for $\ha$ EW values less than -1.0 $\angstrom$, there is a distinct region where the flare rate increases with increasing $\ha$ emission. Because of this, we have focused on exploring the relationship between $\ha$ and the flare rate.

Previous studies have found a saturated relationship between various activity indicators and rotation \citep{Wright2018,Douglas2014,Newton2017}. Instead in \citetalias{Medina2020}, we found that a model where the flare rate in the saturated regime increases modestly with increased $\ha$ emission was favored over a model where the flare rate showed a constant value in the saturated regime. We examined this particular relationship further using the piece-wise function presented in \citetalias{Medina2020} to fit the relationship between $\ha$ and R$_{31.5}$,

\begin{equation}\label{eq:PW_3}
\rm log~R_{31.5}(x_{\ha}) = 
        \begin{cases}
            \gamma (x_{\ha} - X_{\ha_c}) + C &x_{\ha} \leq X_{\ha_{c}}\\
            \beta  (x_{\ha} - X_{\ha_c}) + C &x_{\ha} > X_{\ha_{c}}\\
        \end{cases}
\end{equation}

\noindent where $X_{\ha_{c}}$ is the critical value, C is the rate at which the two regimes intersect, $x_{\ha}$ is the $\ha$ EW, $\gamma$ is the slope for EW values below the critical value, and $\beta$ is the slope above the critical EW value. As we find that there is significant overlap in values of the flare rate for a given $\ha$ EW value between -1.5 and 0.0 $\angstrom$, we fit this function twice at two critical values  $X_{\ha_{c1}}$ and  $X_{\ha_{c2}}$ to reflect this region.  We determine the value of $X_{\ha_{c1}}$ using $\chi^2$ minimization by fitting equation \ref{eq:PW_3} to each $\ha$ EW value and taking the $\ha$ corresponding to the minimum $\chi^2$ as $X_{\ha_{c1}}$. We found that $X_{\ha_{c1}}$ = -0.183 $\angstrom$. We decided to place $X_{\ha_{c2}}$ at -1.5 $\angstrom$ as this is where the lower limit appears to fall.

\begin{figure}[ht]
\includegraphics[scale=.35,angle=0]{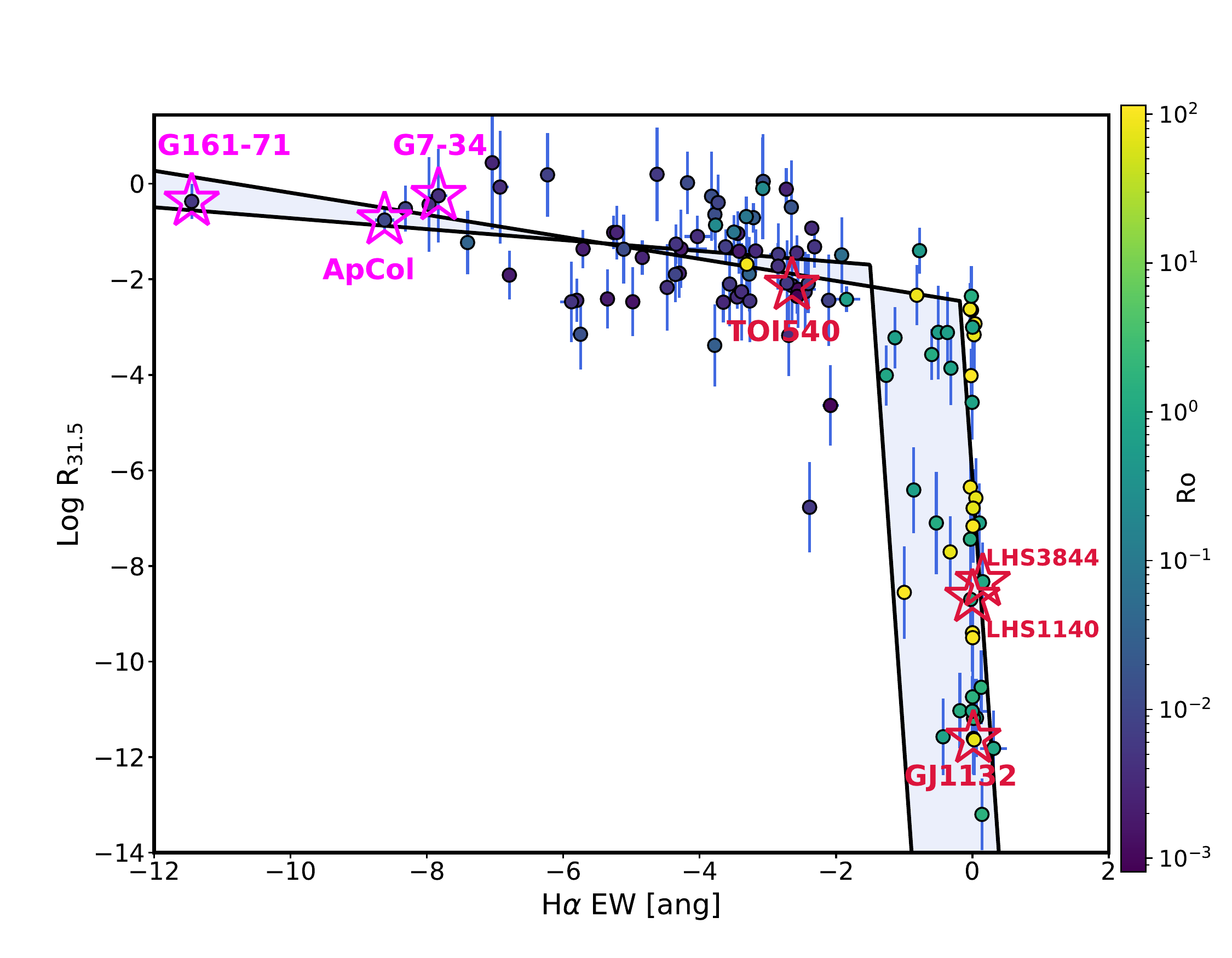}
\caption{The natural log rate of flares per day with energies above E = 3.16$\times$ 10$^{31}$ ergs as a function of measured equivalent widths of $\ha$ for the all-sky sample with \numflares~or more flares and a measured rotation period. The color of the points indicates the value of the Rossby number shown by the colorbar. The black lines show the best fit models using Equation \ref{eq:PW_3} for values of X$_{\ha_{c1}}$ and X$_{\ha_{c2}}$. The shaded area denotes the region between the two fits. Red stars denote known transiting planet hosts. Magenta stars denote members of young moving groups discussed in section \ref{sec:ymg} \label{fig:ha_fit}}
\end{figure}

Using these critical values for $X_{\ha_{c1}}$ and $X_{\ha_{c2}}$, we used \textit{emcee} to determine the values $\gamma$, $\beta$, and C for each critical value labeled with subscripts 1 and 2 respectively. We place uniform priors on each of these parameters.  For $X_{\ha_{c1}}$,  we found the value for C$_1$ is -2.58 $\pm$ 0.23, $\gamma_1$ is -0.26 $\pm$ 0.13, and $\beta_1$ is -19.85 $\pm$ 1.16. For $X_{\ha_{c2}}$ we fix $\beta_2$ = $\beta_1$ and determine the values of C$_2$ = -1.69 $\pm$ 0.09  and $\gamma_2$ = -0.11 $\pm$ 0.03. The fits are shown Figure \ref{fig:ha_fit}.

\subsection{Fraction of Active Fully Convective M dwarfs}\label{sec:active_frac}
\begin{figure}[ht]
\includegraphics[scale=.5,angle=0]{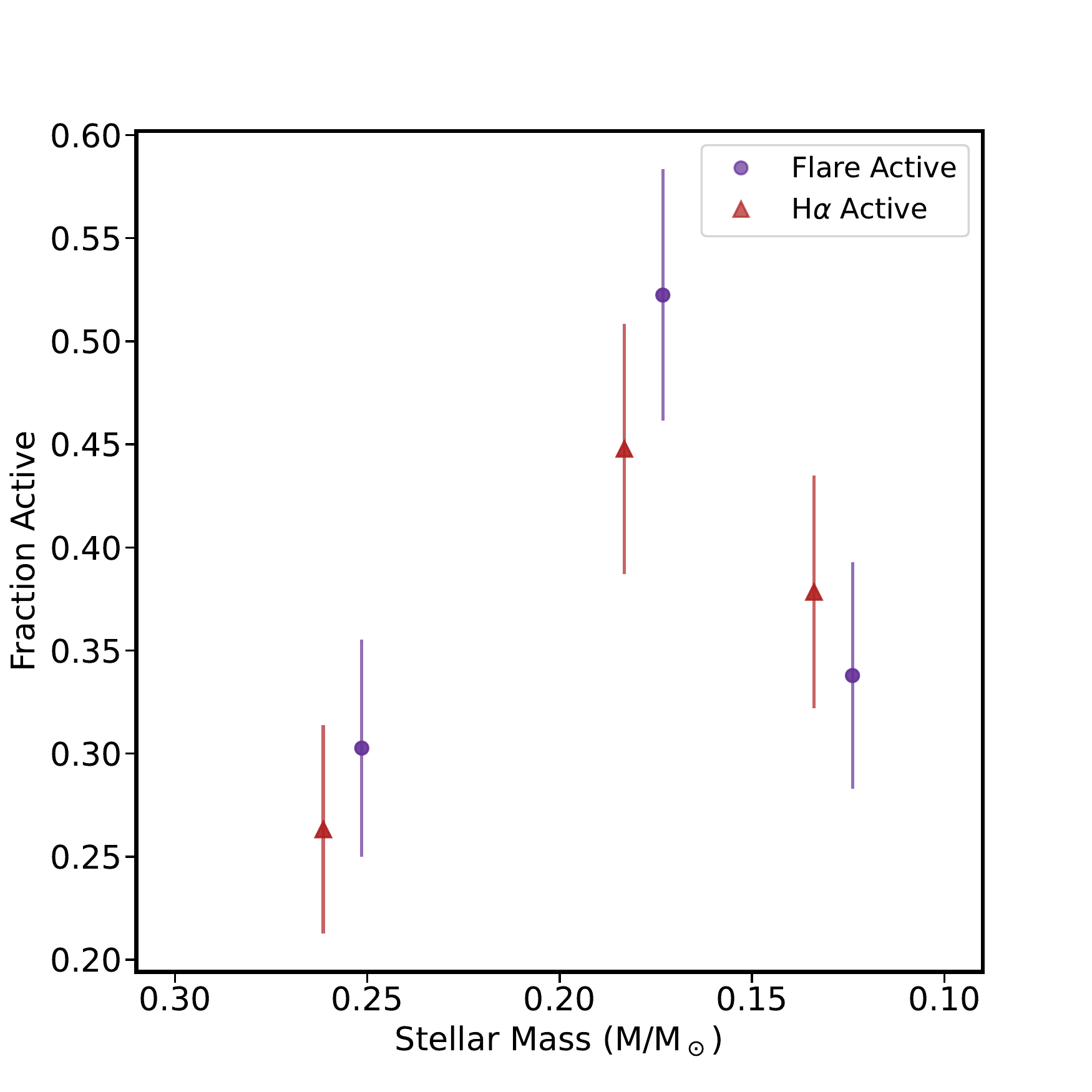}
\caption{The fraction of active stars as defined by Log R$_{31.5}$ $\geq$ to -5 (purple points) and as defined by $\ha$ EW $<$ -1.0 $\angstrom$ (maroon triangles) as a function of stellar mass. The purple and maroon points are offset in stellar mass for clarity. \label{fig:frac_active}} 
\end{figure}

We explore the dependence of magnetic activity on stellar mass. To do this we examine the fraction of stars that are magnetically active in three mass bins (0.1 $<$ M $\leq$ 0.15), (0.15 $<$ M $\leq$ 0.20), and (0.20 $<$ M $\leq$ 0.30) where each mass bin has 73, 68, and 78 stars respectively. We define a star being magnetically active if it has a flare rate, Log R$_{31.5}$ $\geq$ -5. We find the fraction of active stars is 33 $\pm$ 5\%, 53 $\pm$ 6\%, and 31 $\pm$ 5\% for the (0.1 $<$ M $\leq$ 0.15), (0.15 $<$ M $\leq$ 0.20), and (0.20 $<$ M $\leq$ 0.30) mass bins respectively. In addition, we also determine the fraction of stars that are active as defined by having an $\ha$ EW $<$ -1.0 $\angstrom$. We find the fraction of $\ha$ active stars is 36 $\pm$ 5\%, 42 $\pm$ 6\%, and 25 $\pm$ 5\% for the (0.1 $<$ M $\leq$ 0.15), (0.15 $<$ M $\leq$ 0.20), and (0.20 $<$ M $\leq$ 0.30) mass bins respectively. In Figure \ref{fig:frac_active}, we show the fraction of active stars as defined by the flare rate and $\ha$ EW in each mass bin. The error bars were computed using binomial statistics.  We find that the fraction of active stars defined by the flare rate peaks at intermediate mass ranging from 0.15 - 0.20 $\msol$. When considering the $\ha$ indicator, we find that the activity fraction increases from the highest mass bin to the intermediate, but we find that the there is not a statistically significant decrease from the intermediate mass bin to the lowest mass bin. We find the behavior between the fraction of active stars as defined by $\ha$ as a function of stellar mass to be in agreement with previous studies \citep[e.g][]{Hawley1996, Gizis2000, West2004, West2015} where the fraction of field stars that are active as defined by having $\ha$ EW $<$ -1 $\angstrom$ increases with decreasing stellar mass.  However, the fact that the fraction of stars that are in the saturated flare regime does not increase monotonically with decreasing mass is worthy of note.

\subsection{Kinematics in the Galactic Disk}
The Galactic disk contains three populations of stars showing different kinematic properties; the thin, thick disk, and halo \citep[see, e.g.,][]{Bensby2003}. Compared to the thin disk, stars in the thick disk show larger velocity dispersion, an increased scale height, and come from an older population. The halo is generally made up of the oldest stars in the Galaxy with high velocity dispersions and a spherical spatial distribution. \citet{Fantin2019} finds that the thin disk has had approximately constant stellar formation for the past 8 Gyr, the thick disk had an epoch of stellar formation that peaked 10 Gyr ago, and the halo had its peak star formation epoch approximately 11 billion years ago. In this study we determine, the ratio of the probability a given star is a member thick disk over the probability it is a member of the halo,  and the probability a star is in the thick disk over the probability it is in the thin disk. \citet{Newton2016} showed that fully convective M dwarfs with rotation periods less than 10 days were more likely to be dynamically cold thin disk members than their intermediate or slowly  rotating counterparts. We followed the methods described in \citet{Bensby2003} to extend the \citet{Newton2016} study to include updated kinematic information and flare rates. 

\begin{figure*}
\includegraphics[scale=.61,angle=0]{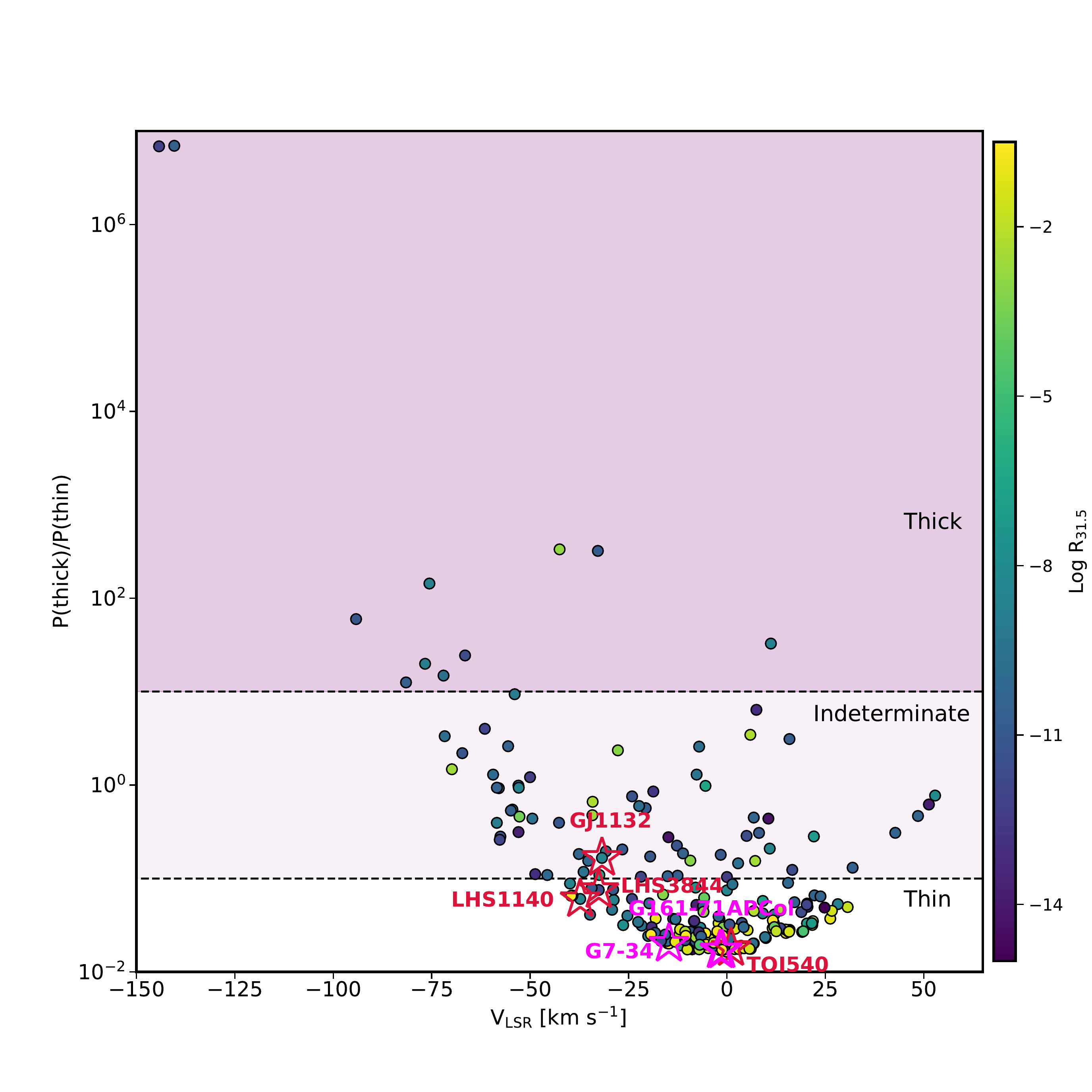}
\caption{The probability $P$ that a star resides in the thick disk, $P(thick)$ divided by the probability that a star resides in the thin disk $P(thin)$ as a function of the V$_{lsr}$. The dashed lines denote the thin disk, P($thick$)/P($thin$) $\leq$ 0.1, the indeterminate region defined as 0.1 $<$ P($thick$)/P($thin$) $<$ 10, and the  thick disk defined as P($thick$)/P($thin$) $\geq$ 10. The color of the points denotes the flare rate, Log R$_{31.5}$, value given by the color bar. Red stars denote known transiting planet hosts in our sample. Magenta stars denote members of young moving groups discussed in section \ref{sec:ymg}.    \label{fig:disk_memb}}
\end{figure*}

\citet{Bensby2003} estimated disk membership based on the  U, V, and W space velocities of each star, the characteristic velocity dispersions $\sigma_{U_{lsr}}$, $\sigma_{V_{lsr}}$, $\sigma_{W_{lsr}}$, asymmetric drift V$_{asym}$ in each component, and the number density of stars present in each component. We used the values presented in Table 1 of \citet{Fantin2019} for  $\sigma_{U_{lsr}}$ = (33, 15, 15) km s$^{-1}$, $\sigma_{V_{lsr}}$ = (40, 32, 28) km s$^{-1}$, $\sigma_{W_{lsr}}$ = (131,106, 85) km s$^{-1}$ and asymmetric drift V$_{asym}$ = (-12, -85, -226) km s$^{-1}$ for the (thin disk, thick disk, halo) respectively. We assumed the local number density of stars to be 83.0\% for the thin disk, 17.0\% for the thick disk, and that 0.1\% of stars reside in the halo of the galaxy \citep{Fantin2019}.  In Figure \ref{fig:disk_memb}, we show the probability $P$ of a star being a thick disk $P(thick)$ member divided by the probability of it being a member of the thin disk $P(thin)$ as a function of V$_{lsr}$ with the color bar indicating the flare rate. We defined the thick disk as P($thick$)/P($thin$) $\geq$ 10, stars that are indeterminate are defined as 0.1 $<$ P($thick$)/P($thin$) $<$ 10, and stars located in the thin disk as P($thick$)/P($thin$) $\leq$ 0.1. We also compute P($thick$)/P($halo$), where P($thick$)/P($halo$) $>$ 10 indicates a high probability a star is a member of the thick disk. We found that with the exception of two stars LHS 288 and GJ 299 where P($thick$)/P($halo$) = 7.5 and 7.9 respectively,  P($thick$)/P($halo$) $>$ 100. LHS 288 and GJ 299 are the outliers at P($thick$)/P($thin$) $>$ 10$^{6}$ in Figure \ref{fig:disk_memb}.  We found that 5.05\% of stars reside in the thick disk, 30.73\% in the indeterminate region, and the remaining 64.22\% reside in the thin disk. 

\begin{figure*}
\includegraphics[scale=.61,angle=0]{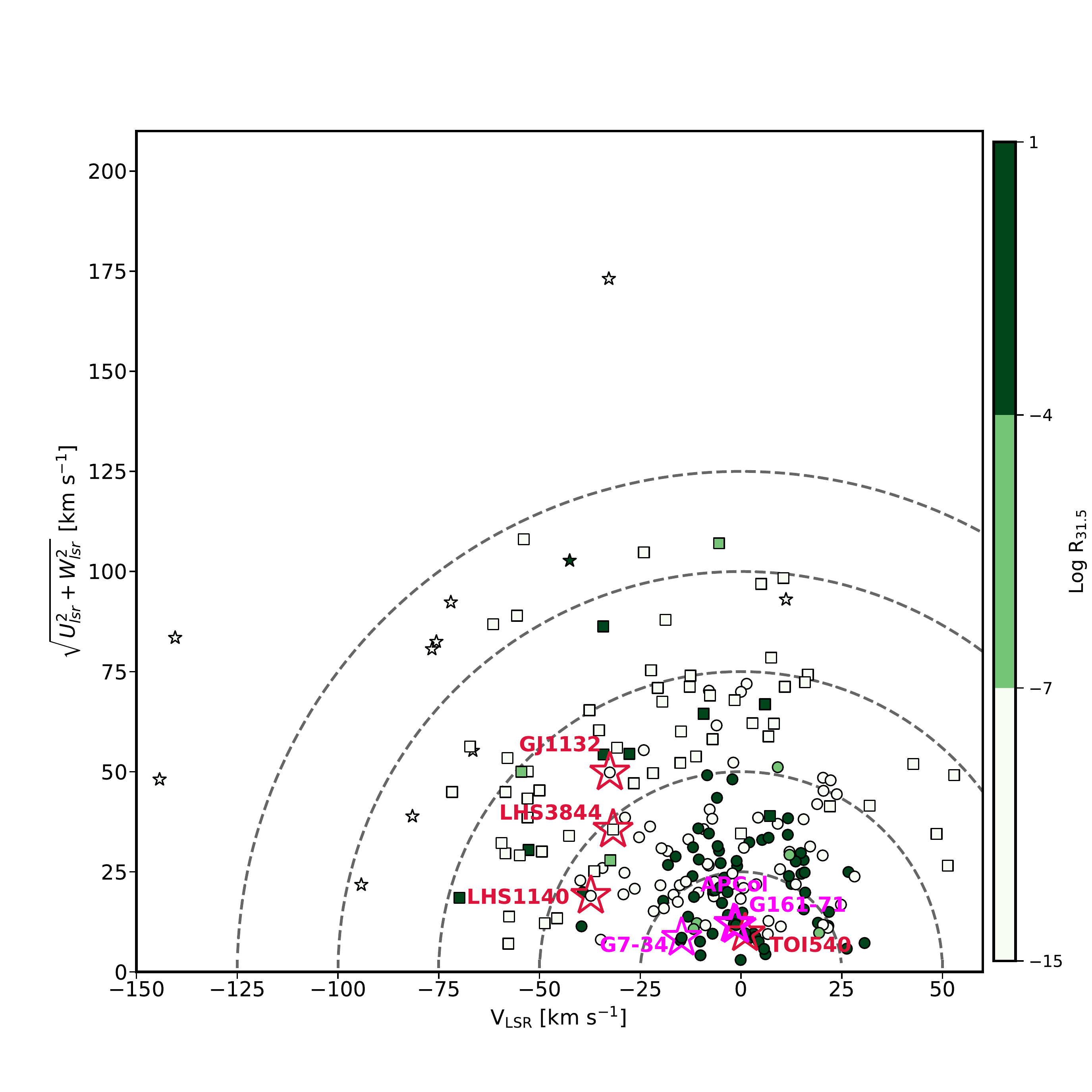}
\caption{The Toomre Diagram which is the defined as $\sqrt{U_{lsr}^2 + W_{lsr}^2}$ as a function of V$_{lsr}$ for all the stars in our sample. The circles represent stars that are likely members of the thin disk, squares represent stars for which the assignment to the thin or thick disk is indeterminate, and star-shaped symbols represent stars that are likely members of the thick disk. The color of the points denotes the flare rate, Log R$_{31.5}$, value given by the color bar. Where we have broken the colorbar up in discrete regions of the Log R$_{31.5}$ =  greater than -4, between -4 and -7, and less than -7.  Lines of constant V$_{tot}$ are indicated with the dashed grey lines and are in steps of 25 km s$^{-1}$. Red stars denote known transiting planet hosts. Magenta stars denote members of young moving groups discussed in section \ref{sec:ymg}.    \label{fig:toomre_dia}}
\end{figure*}

In Figure \ref{fig:toomre_dia} we show the Toomre Diagram which is used to distinguish thin, thick, and halo populations. We see that thin disk stars, and stars for which the assignment is indeterminate with a value P($thick$)/P($thin$) $<$ 1 are all contained within the V$_{tot}$ = 75 km s$^{-1}$ constant velocity contour. Furthermore thin disk magnetically active stars with Log R$_{31.5}$ $>$ -4 are all confined within the V$_{tot}$ = 50 km s$^{-1}$ constant velocity contour indicating that they are kinematically cold. We find that indeterminate stars with 0.1 $<$ P($thick$)/P($thin$) $<$ 10 reside in between V$_{tot}$ = 25--125 km s$^{-1}$. The thick disk stars, P($thick$)/P($thin$) $>$ 10 all reside outside of the V$_{tot}$ = 75 km s$^{-1}$ indicating that this population of stars is kinetically warmer. 

\begin{figure*}
\centering
\includegraphics[scale=.8,angle=0]{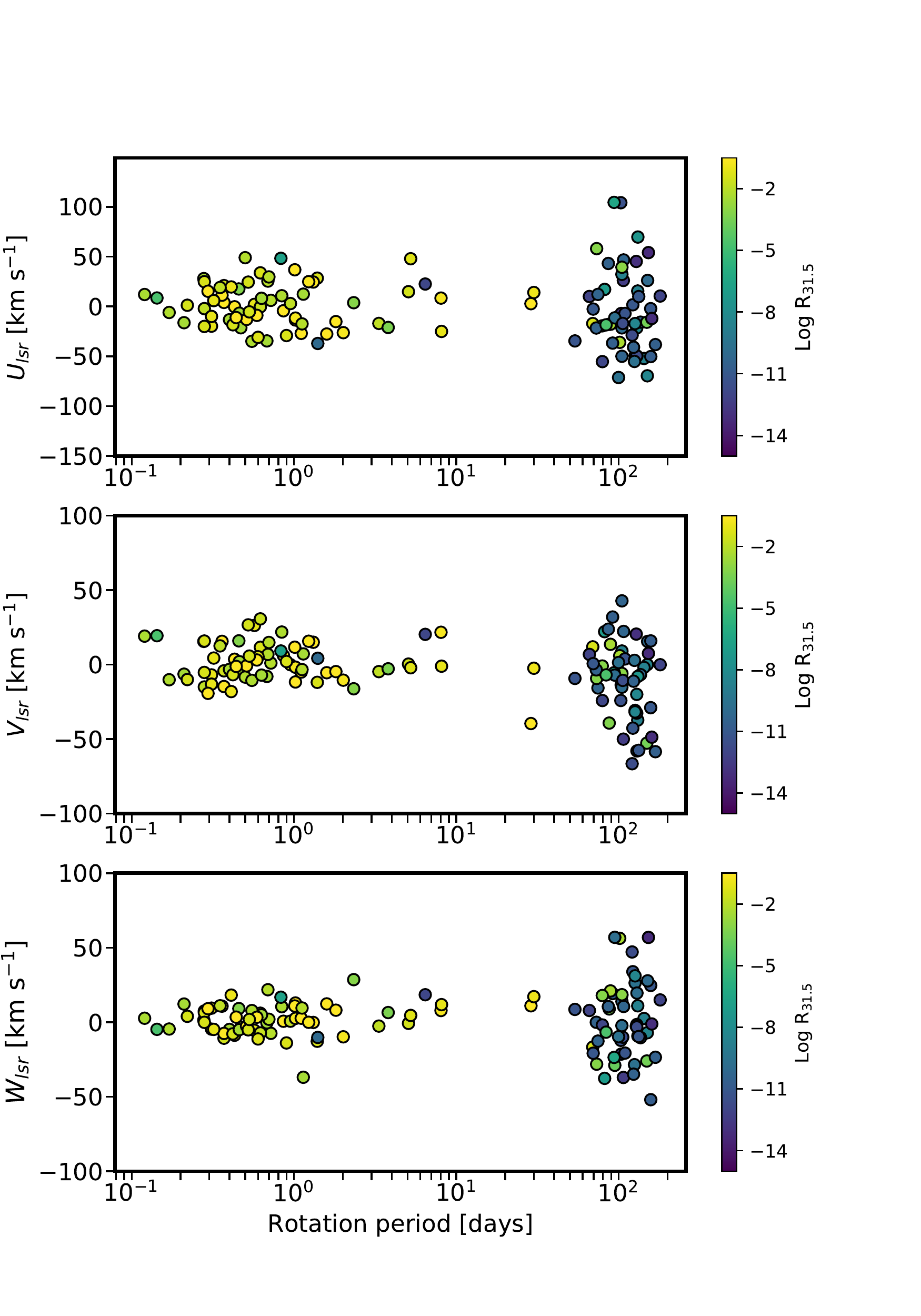}
\caption{The top panel shows the U$_{lsr}$ component of the velocity, the middle panel shows the V$_{lsr}$ component of the velocity, and the bottom panel shows the W$_{lsr}$ component of the velocity all as a function of stellar rotation period. The color of the points denotes the flare rate, Log R$_{31.5}$, given by the color bar.   \label{fig:uvw_flare}}
\end{figure*}

\begin{figure}[ht]
\includegraphics[scale=.35,angle=0]{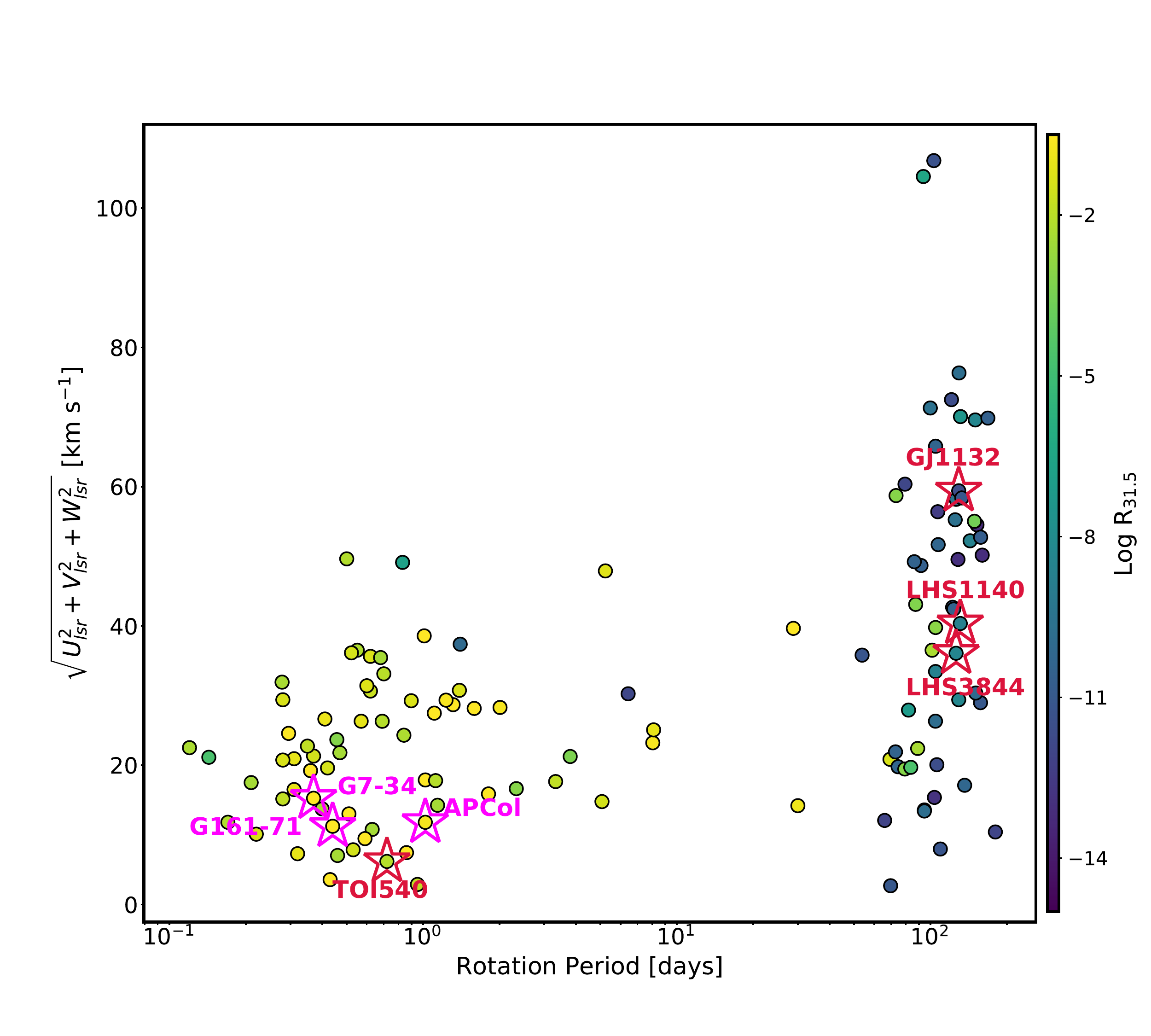}
\caption{The total space velocity, V$_{tot}$ as a function of stellar rotation period. The color of the points denotes the flare rate,Log R$_{31.5}$, given by the color bar. Red stars denote known transiting planet hosts. Magenta stars denote members of young moving groups discussed in section \ref{sec:ymg}. \label{fig:tot_vel}}
\end{figure}

\citet{Newton2016} studied a similar sample of stars in the pre-Gaia era and found that the velocity dispersion of the individual velocity components, (U$_{lsr}$,V$_{lsr}$,W$_{lsr}$), and the total space velocity increased as a function of increasing rotation period. We show the individual (U$_{lsr}$,V$_{lsr}$,W$_{lsr}$) space motions as a function of rotation period with a colorbar indicating the flare rate in Figure \ref{fig:uvw_flare}. We found that in general stars with rotation periods exceeding 90 days show a higher velocity dispersion in each component and have log R$_{31.5}$ that can be 6 orders of magnitude or more below the saturated value than stars with rotation periods $<$ 90 days. At longer rotation periods, we also see evidence of asymmetric drift in the V component where the velocities become increasingly negative at rotation periods longer than 90 days \citep{Eggen1989, West2015, Newton2016}. In Figure \ref{fig:tot_vel} we plot the total space motion as a function of stellar rotation. Here again, we observed that stars with rotation periods that exceed 90 days have low flare rates and a higher total space velocity dispersion.

Previous studies have used a relationship between age and velocity dispersion \citep{Wielen1977,Schmidt2007,Reiners2009b, Faherty2009, Aumer2009,Aumer2016,Newton2016,Yu2018,Kiman2019,Lu2021} to estimate ages of stars in the solar neighborhood. We followed the methods first outlined in \citet{Yu2018} and described further in \citet{Lu2021}, which uses the velocity dispersion in the Z direction $\sigma_{v_z}$, (towards the galactic north pole) as defined by the Galactocentric coordinate system to determine ages. We use \textit{Galpy} to transform heliocentric UVW coordinates into the Galactocentric coordinate system \citep{Bovy2015}.  We calculate $\sigma_{v_z}$ using the median absolute deviation (MAD) multiplied by 1.48 to account for fact that we assume the underlying distribution of Z component velocities is Gaussian. To compute the uncertainties in $\sigma_{v_z}$, we perturbed the Z component of the velocity using 1000 Gaussian deviates with the standard deviation set to the error bar of the Z velocity component value. We broke our sample into bins based on rotation period. A rapidly rotating bin with stars showing P$_{rot}$ $<$ 1 day, a moderately rapidly rotating bin 1 $<$ P$_{rot}$ $<$ 10 days, an intermediate rotation period bin 10 $<$ P$_{rot}$ $<$ 90 days, and a slowly rotating bin with P$_{rot}$ $>$ 90 days. We computed $\sigma_{v_z}$
for each bin finding that $\sigma_{v_z}$ = 8.63 $\pm$ 1.52 km~s$^{-1}$, 9.42 $\pm$ 2.99 km~s$^{-1}$, 17.56 $\pm$ 4.97 km~s$^{-1}$, and 29.71 $\pm$ 4.93 km~s$^{-1}$, respectively.

\citet{Lu2021} provides the following equation to compute the age based on $\sigma_{v_z}$, 

\begin{equation}\label{eq:young_ages}
    ln~(age) = \beta~{\rm ln}~\sigma_{v_z} + \alpha 
\end{equation}

\noindent where $\beta$ = 1.58 $\pm$ 0.19 and $\alpha$ = -2.80 $\pm$ 0.53. As noted in \citet{Lu2021}, \citet{Yu2018} computed the age velocity relation (AVR) for a sample of metal-poor thick disk stars and a sample of metal-rich thin disk stars, but did not provide the value of the intercept for the metal-rich thin disk sample. As such \citet{Lu2021} re-estimated equation \ref{eq:young_ages} and this is the one we use in this study.

 We found that the rapidly rotating bin has an age of 1.8 $\pm$ 0.5 Gyr and  mean log R$_{31.5}$ = -1.69 $\pm$ 0.07, the moderately rotating bin has an age of 2.1 $\pm$ 1.1 Gyr and mean log R$_{31.5}$ = -2.11$\pm$ 0.12 , the intermediate rotators have an age of 5.6 $\pm$ 2.7 Gyr and mean  log R$_{31.5}$ = -6.54 $\pm$ 0.15, and the slow rotators have ages of 12.9 $\pm$ 3.5 Gyr and mean log R$_{31.5}$ = -9.65 $\pm$ 0.12. We computed the uncertainty in the age by repeating the age calculation for each $\sigma_{v_z}$ value plus and minus its uncertainty. The AVR presented in \citet{Yu2018} is not the only relation presented in the literature. \citet{Aumer2009} and\citet{Aumer2016} present AVRs that also use only the vertical velocity dispersion and a power law with a different functional form than \citet{Yu2018} to describe the relationship between age and velocity dispersion.  \citet{Wielen1977} presents a relation using the total velocity dispersion $\sigma_{tot} = \sqrt{\sigma_u^2 + \sigma_v^2 + \sigma_w^2}$. We determined ages using the methods described in \citet{Wielen1977} and \cite{Aumer2016} finding consistency between the ages we computed and those computed using other methods. We chose to use the AVR presented in \citet{Yu2018} as it incorporates updated kinematic information from the Gaia mission.  In Table \ref{tab:age_vel}, we present the bins, number of stars in each bin, the mean period, $\sigma_{v_z}$, and their assigned ages.

For completeness we also determined $\sigma_{v_z}$ and estimate an age for stars with no measured rotation period. We found that $\sigma_{v_z}$ = 26.01 $\pm$ 3.01 km~s$^{-1}$ leading to an age of 10.5 $\pm$ 1.9 Gyr. The age and $\sigma_{v_z}$ of these stars is consistent with the age of stars with rotation periods exceeding 90 days. Thus we postulate that the majority of the stars without a measured rotation period are old and rotate slowly. This statement is further supported by the fact that all of these stars do not show $\ha$ in emission.
 
Although we broke up our sample into four distinct rotation period bins, $\sigma_{v_z}$ for the stars with rotation periods $<$ 1 day and $\sigma_{v_z}$ for stars with rotation periods between 1 and 10 days are statistically indistinguishable and thus their inferred ages are indistinguishable. As such, we combined the rapidly rotating and moderately rotating bins and determined an age of 2.0 $\pm$ 1.2 Gyr.  With this, we posit that stars can reside in the very active phase associated with the saturated regime for at least 2 Gyr.

\subsection{$\sigma_{v_z}$ as a function of Stellar Mass}\label{sec:sig_z}
\begin{figure}[ht]
\includegraphics[scale=.60,angle=0]{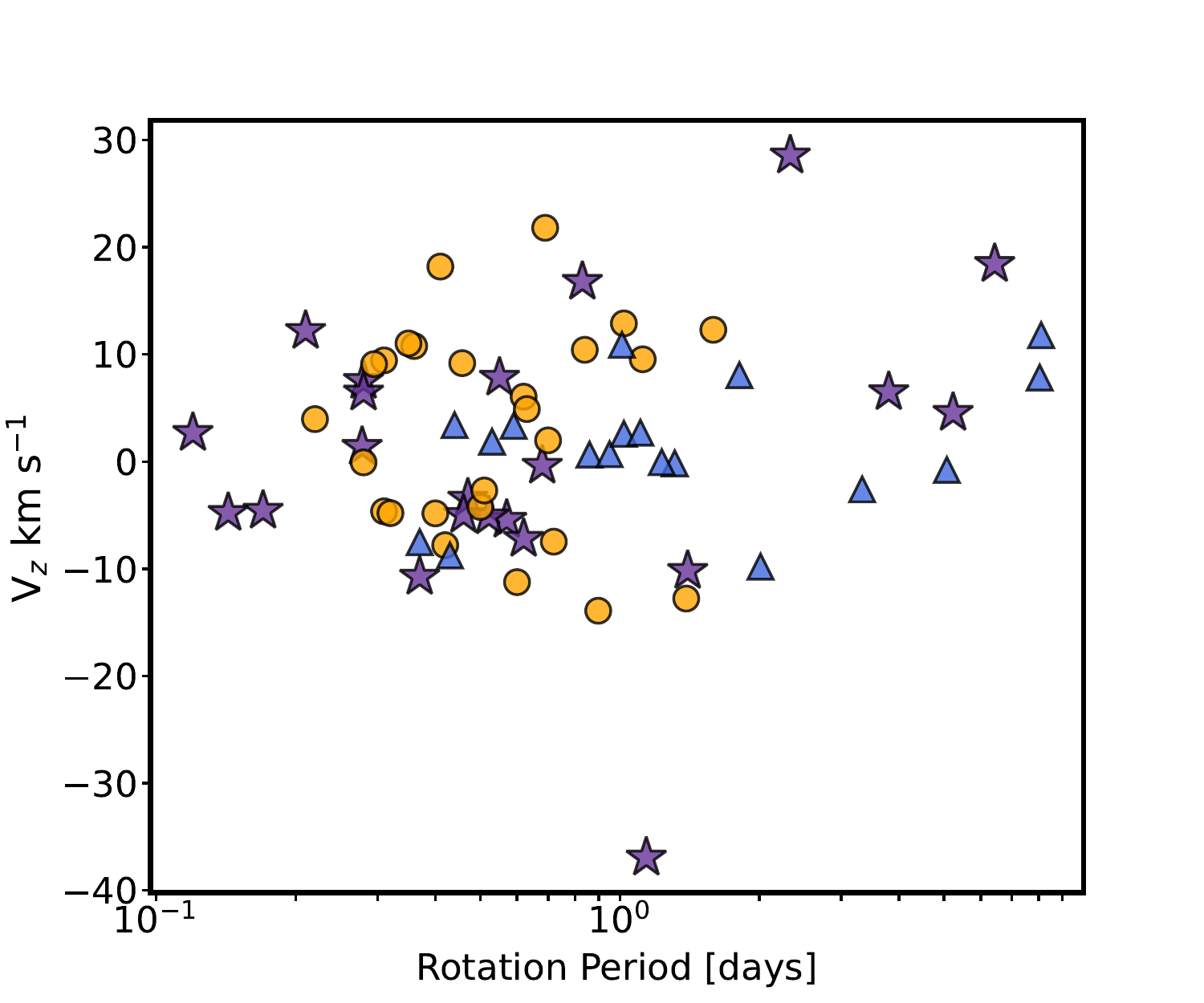}
\caption{V$_z$ as a function rotation period for stars with p${\rm rot} < $ 10 days. The purple stars have masses ranging from [0.1--0.15), the orange circles have masses ranging from [0.15--0.2), and the purple triangles have masses ranging from [0.2, 0.3].  \label{fig:sig_z}}
\end{figure}

We calculated the vertical velocity dispersion as a function of rotation period for the three mass bins presented in Section \ref{sec:active_frac} for stars with rotation periods less than 10 days. We implemented the 10 day rotation period cut-off because we were interested in examining the relationship between age and stellar mass in the saturated regime in particular.  We found that $\sigma_{v_z}$ = 8.83 $\pm$ 2.34 km~s$^{-1}$, 11.09 $\pm$ 2.60 km~s$^{-1}$, and 3.11 $\pm$ 1.30 km~s$^{-1}$ for the (0.1 $<$ M $\leq$ 0.15), (0.15 $<$ M $\leq$ 0.20), and (0.20 $<$ M $\leq$ 0.30) mass bins respectively. We then determined the ages of each respective mass bin using the relations from \citet{Lu2021}. We found ages of 1.9 $\pm$ 0.9 Gyr, 2.7 $\pm$ 1.0 Gyr, and 0.6 $\pm$ 0.3 Gyr for the low, intermediate, and high mass bins respectively. From this,  we observe that the stars in our lowest and intermediate mass bin are older than stars in our highest mass bin implying that lower mass stars stay in the saturated regime of magnetic on average longer than high mass stars. In Figure \ref{fig:sig_z}, we present the V$_z$ component of the velocity as a function of rotation period for the low, intermediate, and high mass bins.

\begin{center}
\begin{deluxetable*}{lcccccc} \label{tab:age_vel}
\tablecaption{Average Ages, Flare Rates, and Velocity Dispersions}
\tablehead{ 
\colhead{Period Bin} & 
\colhead{N Stars} &
\colhead{Mean P} &   
\colhead{$\sigma_{v_z}$} &
\colhead{Log R$_{31.5}$} &
\colhead{Est. Age} \\
\colhead{[days]} &
\nocolhead{yup}  & 
\colhead{[days]} &
\colhead{[km~s$^{-1}$]} &
\colhead{[log flares day$^{-1}$]} &
\colhead{[Gyr]}}
\startdata
P $<$ 1  & 45  & 0.5 & 8.63 &  -1.69 & 1.8 $\pm$ 0.5\\
1 $<$ P $<$ 10 & 21 & 2.8 & 9.42  & -2.11 & 2.1 $\pm$ 1.1\\
10 $<$ P $<$ 90 & 16 & 70.3  & 17.56  & -6.54 & 5.6 $\pm$ 2.7\\
P $>$ 90& 40   & 124.7 & 29.71 & -9.65 & 12.9 $\pm$ 3.5\\
\enddata
\end{deluxetable*}
\end{center}

\subsection{Membership in Young Associations}\label{sec:ymg}
We used BANYAN $\Sigma$ \citep{Gagne2018} to determine whether any stars in our all-sky sample are members of young stellar associations. The BANYAN $\Sigma$ tool takes as inputs the coordinates, radial velocity, proper motion in right ascension, $\mu_{\alpha}$, and declination, $\mu_{\delta}$ and parallax of the star. We found the majority of our stars (216) to have the highest probability of being field stars. However,  we found that three stars have probabilities greater than 99.99\% of being associated with young moving groups. G 7-34 has a 99.997 \% probability of being a member of the AB Doradus moving group that has an age of 149 Myr and has already been noted in \citet{Bell2015} as being a member of this moving group. AP Col and G 161-71 have probabilities of 99.996\% and 99.994\% respectively of being members of the Argus Association which has an age of 40 Myr \citep{Torres2008,Zuckerman2019}. AP Col has been noted as being a pre-main sequence star and being a likely member of the Argus Association by \citet{Riedel2011} and G 161-71 was previously noted as being a member of Argus by \citet{Bartlett2017}. G 161-71 has the greatest amount of $\ha$ in emission with an EW = -11.45 $\angstrom$ of all stars in sample. AP Col shows the second highest $\ha$ EW = -8.61 $\angstrom$, and G 7-34 shows the 5th highest with an $\ha$ EW = -7.82 $\angstrom$. All of these stars have flare rates and rotation periods that place them in the saturated regime of Figure \ref{fig:prot_fl}. 




\section{Discussion and Conclusions}\label{sec:DC}
We used  high-resolution spectroscopic data to measure chromospheric magnetic activity indicators, photometry to measure flare rates and rotation periods, and galactic velocities to constrain the ages of a volume complete sample of 219 fully convective M dwarfs with masses between 0.1 and 0.3$\msol$ that reside within 15 parsecs. The flare rates, chromospheric activity indicators, and rotation periods for 125 of these stars were initially presented in \citetalias{Medina2020}. In this study, we add 94 additional stars that were observed in the northern ecliptic hemisphere during the second year of the primary mission of \textit{\textit{TESS}}. We measured seven new rotation periods with \textit{\textit{TESS}} photometry that ranged from 0.29 to 1.31 days and eight new rotation periods with MEarth North Photometry ranging from 70 to 160 days.  We found that fully convective M dwarfs fall largely into two populations: 29\% have $\ha$ in emission, a saturated flare rate equal to log R$_{31.5}$ = -1.32 $\pm$ 0.06, have Rossby numbers less than 0.50 and form a kinematically cold population with an average age of 2.0 $\pm$ 1.2 Gyr. The remaining 71\% show little to no $\ha$ in emission, have log R$_{31.5}$ $\leq$ -3.00 $\pm$ 0.67, rotation periods, when measured, that exceed exceed 90 days, and have a large galactic velocity dispersion with a mean age of 12.9 $\pm$ 3.5 Gyr. For 99 of the 152 stars in the second group, the photometric rotation period has not been determined, we expect all of these stars to have rotation periods greater of 90 days based on their activity level and kinematics. 

We found that the stars in our all-sky sample show a common power-law exponent for the flare frequency distribution of $\alpha$ =  1.984 $\pm$ 0.019. This is consistent with the value of $\alpha$ $\approx$ 2 that has been found in previous spaced based studies using Kepler, K2, or \textit{\textit{TESS}} for smaller samples of fully convective M dwarfs \citep[][see \citetalias{Medina2020} for an overview of those studies]{Hawley2014,Silverberg2016,Davenport2020,Ilin2019,Ilin2021} with our values being more precise than in previous works. However, \citet{Feinstein2020} in a study of stars with temperatures below 4000 Kelvin in young moving groups with ages up to 750 Myr found that $\alpha$ = 1.58 $\pm$ 0.03. This disagreement brings up the question whether $\alpha$ may change as a function of stellar age. For the three stars that are probable members of young moving groups with ages less than 150 Myr: G 7-34, G 161-71, and AP Col we found $\alpha$ = 1.90 $\pm$ 0.12, 1.86 $\pm$ 0.10, and 1.89 $\pm$ 0.13. \citet{Ilin2019} and \citet{Ilin2021} in studies of flare stars in different clusters with known ages observed by K2 found of that $\alpha$ appears to be approximately 2 with no dependence on age or spectral type. We found using the methodology and software \textit{STELLA} presented in \citet{Feinstein2020} on G 7-34, G 161-71, and  AP Col that many spurious lower-energy flares were being detected. The detection of these spurious signals could account for the discrepant value of $\alpha$ that is obtained in that study. Furthermore,  we found no indication of a change in $\alpha$ for stars with different kinematic ages. An $\alpha$ value of 2 or greater means means that low-energy flares occur frequently enough that the energy they deposit to the stellar corona may account for observed coronal temperatures $\approx$ 10$^6$ Kelvin. Furthermore, this common value of $\alpha$ points to a common mechanism for the production of flares on fully convective M dwarfs and perhaps all stars with convective envelopes.


We compute the mean ages for groups of stars in different rotation period bins. We find that stars with rotation periods less than 10 day have ages of 2.0 $\pm$ 1.2 Gyr, stars with 10 $<$ P $<$ 90 days have ages of 5.6 $\pm$ 2.7 Gyr, and stars with rotation periods greater than 90 days have an age of 12.9 $\pm$ 3.5 Gyr.  We find that stars with a saturated flare rate and Rossby number less than 0.5 days likely belong to a younger, kinematically colder population. Through our kinematic analysis, we found that two stars in our sample G 161-71 and AP Col are members of the young stellar association Argus with an age of 40 Myrs and one star, G 7-34 is a member of the AB Doradus Moving Group with an age of 149 Myrs. We found that these stars are distinct in that they have some of the highest $\ha$ emission with EWs of -8.61 $\angstrom$, -11.45 $\angstrom$, and -7.82 $\angstrom$ for AP Col, G 161-71, and G 7-34 respectively. \citet{Medina2022} recently showed that G 7-34 differed from the other 9 active stars in that sample in that its $\ha$ emission varied in phase with the stellar rotation period of the star. This in phase variation indicated that the dominant source of $\ha$ variability on G 7-34 originates from constant emission from fixed magnetic structures, such as starspots and plage. This is in contrast to the other 9 active, but older stars in their sample which showed no correlation between $\ha$ variability and rotational phase; the dominant source of $\ha$ variability in these stars is likely due to stellar flares. It is not yet clear why the dominant source of $\ha$ variability on this young star is different than older active stars, but probing the $\ha$ variability timescale of G 161-71 and AP Col may provide insight into how the interplay between emission from magnetic structures and flares evolves with time.   

\subsection{At What Age Do Fully Convective Stars Spin Down?}

Our results, and those of others concerning coronal x-ray emission \citep{Stauffer1987,Pizzolato2003, Wright2018}, $\ha$ emission \citep{Douglas2014, Newton2017}, and magnetic field strength \citep{Reiners2022} show that fully convective M dwarfs fall largely into two categories: rapidly rotating stars that show a saturated magnetic activity that is independent of rotation up to a critical value, and stars that fall into the unsaturated activity-rotation regime. Theoretical studies posit that stars occupying the saturated regime of magnetic activity have complex multipolar magnetic field topologies, few open field lines \citep{Garraffo2018}, and a dynamo that is weakly coupled to the stellar wind \citep{Brown2014}. These theoretical studies suggest that stars occupying the unsaturated regime have dipolar fields, many open field lines, and have a dynamo that is strongly coupled to the stellar wind \citep{Garraffo2018, Brown2014}. A recent observational study by \citet{Reiners2022} found that saturation of the average surface magnetic field is due to a saturation of the magnetic dynamo rather than a saturation of the stellar surface by magnetic features. They suggest that the surface magnetic field generated is limited by the available kinetic energy and thus can increase only until it reaches the kinetic field. It is currently unclear what role, if any, the topology of the field plays. Furthermore, the duration of the saturated phase and at what age the transition from one regime to the other takes place is currently unknown.


Here we estimate, using a very simplified model, on an age that the transition may occur based on our results and the picture put forth by \citet{Brown2014}. As discussed previously,  \citet{Fantin2019} showed that the star formation in the thin disk has been approximately constant for past 8 Gyr with a stark drop off in the star formation rate at older ages. We show that in Figure \ref{fig:disk_memb} that the majority of our stars are members of the thin disk. As such, we assume that stellar formation of the stars in the all-sky sample has been constant for the past 8 Gyr. If we take these assumptions to be true, coupled with the findings that 29\% of stars reside in the saturated regime and 71\% in the unsaturated regime, we posit that the age at which stars spontaneously transition from the saturated to unsaturated regimes occurs at approximately 2.4 $\pm$ 0.3 Gyr. This estimation is consistent with the kinematic ages we measured in this study where we find that stars with rotation periods less than 10 days have ages of 2.0 $\pm$ 1.2 Gyr, stars with 10 $<$ P$_{rot}$ $<$ 90 days have ages around 5.6 $\pm$ 2.7 Gyr, and stars with rotation periods greater than 90 days have an age of 12.9 $\pm$ 3.5 Gyr. The estimate of 2.4 $\pm$ 0.3 Gyr fits in where expected, between the rotation period bin containing stars with rotation periods between 10 and 90 days. We do however, note that this is a lower limit as some of our stars do reside in the thick disk where the stellar formation history has not been constant. We also find that when we examine the mass dependence of the transition for stars in the saturated regime with Prot $<$ 10 day, we find that the average age of stars with Prot $<$ 10 days increases from 0.6 $\pm$ 0.3 Gyr for stars with masses between 0.2--0.3$\msol$  to 2.3 $\pm$ 1.3 Gyr for stars with masses between 0.1--0.2 $\msol$.


With the exception of planets around active M stars like K2-25 b, which is not in our sample,  \citep{Gaidos2020} and TOI-540 b \citep{Ment2021} which is in our sample, the planets we discover largely orbit inactive stars that reside in the unsaturated regime. However if it is true that these stars sustain high levels of extreme ultra-violet and x-ray radiation associated with the saturated regime for typically 2.4 Gyr, this would have a profound effect on the atmospheres of planets orbiting this stars.  These atmospheres are those that will be observable in the near-term with the next generation of space based telescopes and ground based extremely large telescopes. In this study, we characterized the activity in the saturated regime and the age the transition occurs. These constraints are pertinent to the planets orbiting fully convective stars; in order to fully understand the results of upcoming atmospheric studies, the stellar radiation environment and its history must be reconstructed. 


\acknowledgments{We would like to thank the referee for a thoughtful review that improved the manuscript. The authors would like to thank David Latham, Jessica Mink, Gilbert Esquerdo, Perry Berlind, and Michael Calkins for scheduling,  collection, and reduction of the TRES data. This research has used data from the CTIO/SMARTS 1.5m telescope, which is operated as part of the SMARTS Consortium by RECONS (www.recons.org) members Todd Henry, Hodari James, Wei-Chun Jao , and Leonardo Paredes.  AAM is supported by NSF Graduate Research Fellowship Grant No. DGE1745303. AAM would like to thank Charlie Conroy for insightful conversations about the star formation history of the the Milky Way. AAM would like to thank Emily Pass of her expertise on bisector analyses. This work is made possible by a grant from the John Templeton Foundation. The opinions expressed in this publication are those of the authors and do not necessarily reflect the views of the John Templeton Foundation. The MEarth Team gratefully acknowledges funding from the David and Lucile Packard Fellowship for Science and Engineering (awarded to D.C.). This material is based upon work supported by the National Science Foundation under grants AST-0807690, AST-1109468, AST-1004488 (Alan T. Waterman Award), and AST-1616624. This material is based upon work supported by the National Aeronautics and Space Administration under grant under grants 80NSSC19K0635, 80NSSC19K1726, 80NSSC21K0367, 80NSSC22K0165, and 80NSSC22K0296 in support of the TESS Guest Investigator Program and grant 80NSSC18K0476 issued through the XRP program. This paper includes data collected by the \textit{TESS} mission, which are publicly available from the Mikulski Archive for Space Telescopes (MAST). This work has made use of data from the European Space Agency mission Gaia (https://www.cosmos.esa.int/gaia), processed by the Gaia Data Processing and Analysis Consortium (DPAC, https://www.cosmos.esa.int/web/gaia/dpac/consortium). Funding for the DPAC has been provided by national institutions, in particular the institutions participating in the Gaia Multilateral Agreement.   This work has used data products from the Two Micron All Sky Survey, which is a joint project of the University of Massachusetts and the Infrared Processing and Analysis Center at the California Institute of Technology, funded by NASA and NSF.} 

\facilities{\textit{TESS}, MEarth, FLWO:1.5m (TRES), CTIO/SMARTS:1.5m (CHIRON)} 

\software{{\sc celerite} \citep{Foreman-Mackey(2017)}, {\sc exoplanet} \citep{exoplanet:exoplanet}, {\sc PYMC3} \citep{exoplanet:pymc3}, {\sc python}}

This research made use of {\sc exoplanet} \citep{exoplanet:exoplanet} and its dependencies \citep{exoplanet:astropy13, exoplanet:astropy18, exoplanet:exoplanet, exoplanet:foremanmackey17, exoplanet:foremanmackey18, exoplanet:kipping13, exoplanet:luger18, exoplanet:pymc3, exoplanet:theano}.

\clearpage

\bibliographystyle{aasjournal}
\bibliography{references}{}

\end{document}